\documentclass{article}
\usepackage{amsmath,amsfonts,amssymb}
\usepackage[dvips]{graphicx}
\usepackage{here}
\usepackage{ulem}
\usepackage{color}
\newcommand{\nc}{\newcommand}
\nc{\rnc}{\renewcommand}
\nc{\nn}{\nonumber}
\nc{\db}{\displaybreak[0]\\}
\rnc{\[}{\left[}
\rnc{\]}{\right]}
\rnc{\(}{\left(}
\rnc{\)}{\right)}
\nc{\lt}{\left\{}
\nc{\rt}{\right\}}
\nc{\sg}{\sigma}
\nc{\g}{\gamma}
\nc{\lam}{\lambda}
\rnc{\b}{\beta}
\nc{\vp}{\varphi}
\nc{\bra}{\langle}
\nc{\ket}{\rangle}
\rnc{\i}{{\rm i}}
\rnc{\d}{{\rm d}}
\nc\sign{\text{sign}}
\nc{\ds}{\displaystyle}

\nc\sn{\text{sn}}
\nc\cn{\text{cn}}
\nc\dn{\text{dn}}
\nc\am{\text{am}}
\nc\sech{\text{sech}}
\nc\nt{n}
\nc{\cd}{\cdots}
\nc{\sm}[2]{\sum_{#1=1}^{#2}}
\nc\hp{\hat{\psi}}
\nc\hpd{\hat{\psi}^\dagger}
\nc{\mfperi}{\psi_\text{C}}
\nc\qs{\psi_\text{Q}}
\nc{\red}{\textcolor{red}}
\nc{\blue}{\textcolor{blue}}
\nc{\green}{\textcolor{green}}
\nc{\sred}[1]{\textcolor{red}{\sout{#1}}}
\begin{document}
\title{
Quantum states of dark solitons in the 1D Bose gas  
}
%
\author{
  Jun Sato           ${}^1$ \thanks{jsato@jamology.rcast.u-tokyo.ac.jp},   \ \
  Rina Kanamoto ${}^2$ \thanks{kanamoto@meiji.ac.jp}, \ \ 
  Eriko Kaminishi   ${}^3$ \thanks{kaminishi@spin.phys.s.u-tokyo.ac.jp}, \ \ 
 and Tetsuo Deguchi   ${}^4$ \thanks{deguchi@phys.ocha.ac.jp}
\\
\\ 
\it ${}^1$
Research Center for Advanced Science and Technology, University of Tokyo, \\\it 
4-6-1 Komaba, Meguro-ku, Tokyo 153-8904, Japan\\
\it ${}^2$
Department of Physics, Meiji University, 
Kawasaki, Kanagawa 214-8571, Japan\\
\it ${}^3$
Department of Physics, 
University of Tokyo, \\\it 
7-3-1 Hongo, Bunkyo-ku, Tokyo 113-0033, Japan\\
\it ${}^4$
Department of Physics, Faculty of Core Research, 
\\ 
\it 
Ochanomizu University, 2-1-1 Ohtsuka, Bunkyo-ku, Tokyo 112-8610, Japan}
\maketitle
\begin{abstract}
We present a series of quantum states that are characterized by dark solitons 
of the nonlinear Schr\"{o}dinger equation (i.e. the Gross-Pitaevskii equation) 
for the one-dimensional (1D) Bose gas interacting through the repulsive delta-function potentials. The classical solutions satisfy the periodic boundary conditions and we 
simply call them classical dark solitons. Through exact solutions we show corresponding aspects between the states and the solitons in the weak coupling case: the quantum and classical density profiles completely overlap with each other not only at an initial time but also at later times over a long period of time, and they move together with the same speed in time; 
the matrix element of the bosonic field operator 
between the quantum states has exactly the same profiles of the square amplitude and the phase as the classical complex scalar field of a classical dark soliton not only at the initial time but also at later times, and the  corresponding profiles move together for a long period of time. 
We suggest that the corresponding properties hold rigorously in the weak coupling limit. 
Furthermore, we argue that the lifetime of the dark soliton-like density profile 
in the quantum state becomes infinitely long as the coupling constant approaches zero, 
by comparing it with the quantum speed limit time.  
Thus, we call the quantum states {\it quantum dark soliton states}. 
\end{abstract}
%
%
\section{Introduction}
The experimental technique of trapped one-dimensional atomic gases \cite{Ketterle, Esslinger} has now become a fundamental tool for revealing nontrivial effects in quantum many-body  systems \cite{Kinoshita,prethermal}. For the interacting Bose gas in one dimension (1D), the first set of exact results goes back to  the pioneering work of Girardeau \cite{Girardeau} on the impenetrable Bose gas where the strong interacting limit is considered. The 1D Bose gas interacting with the delta-function potentials, i.e. the Lieb-Liniger (LL) model, gives a solvable model for interacting bosons in 1D \cite{Lieb-Liniger}, where it is integrable even when the interaction parameter is generic. For the impenetrable Bose gas which corresponds to the Tonks-Girardeau (TG) limit, i.e. the strong interacting limit of the LL model, the one-body reduced density matrix is derived and successfully expressed in terms of the determinant of a Fredholm operator \cite{Lenard}. The exact result is followed by several important developments in mathematical physics \cite{Jimbo-Miwa, Korepin}. Furthermore, dynamical correlation functions of the LL model are now systematically derived \cite{KKMST2012}. 

Soliton-like localized excitations in a periodic 1D Bose gas have attracted much interest recently and have been studied theoretically \cite{Refa, Carr, Rina}. 
Here we remark that dark solitons are created experimentally in cold atoms \cite{review}, 
for instance, by the phase-imprinting method \cite{Lewenstein,Sengstock} (See also \cite{Oberthaler}). 
Localized quantum states are important for investigating dynamical responses of interacting quantum systems. Quantum dark solitons in confining potentials are studied by semiclassical quantization \cite{Gangardt}, and those in thermal equilibrium of a quasi-1D Bose gas by generating classical field ensembles \cite{PolishSchool}. However, it is not clear even at zero temperature how we can construct quantum states associated with dark solitons in the many-body system of the LL model. 

Let us consider the Gross-Pitaevskii (GP) equation, which describes Bose-Einstein condensation (BEC) in the mean-field approximation \cite{Leggett}. We also call it the nonlinear Schr{\"o}dinger equation. The GP equation has dark soliton solutions for the repulsive interactions, while it has bright soliton solutions for the attractive interactions \cite{soliton}. It was conjectured that dark solitons are closely related to Lieb's type-II excitations, i.e. one-hole excitations, by carefully studying the dispersion relations \cite{Takayama}. The dispersion relations of the LL model are briefly compared with those of the classical nonlinear Schr{\"o}dinger equation in the weak coupling limit \cite{KMF}. 
However, it has not been shown how one can construct such a quantum state that leads to a dark soliton in the classical limit or what kind of physical quantity can show a property of a dark soliton for some quantum state. Here we remark that each of the type-II eigenstates has a flat density profile since the Bethe ansatz eigenstates are translationally invariant. Moreover, we consider dark solitons under the periodic boundary conditions, which are expressed in terms of elliptic functions \cite{Refa,Carr,Rina}.

In this paper we demonstrate that a quantum state derived from the Bethe ansatz eigenvectors of the LL model by taking the Fourier transform of the type-II excitations over a branch \cite{SKKD1} has many aspects closely related to classical dark solitons of the GP equation under the periodic boundary conditions. We call the state the {\it quantum dark soliton state} and a classical dark soliton under the periodic boundary conditions simply a {\it classical dark soliton}.  
Through the Bethe ansatz we show several corresponding aspects in the weak coupling regime. 
 Firstly, the density profile of the quantum dark soliton state at an initial time is consistent with that of a classical dark soliton. Here we call the graph of the expectation value of the local density operator for a given state versus the position coordinate the density profile of the state,  and for a quantum dark soliton state we simply call it the density profile of  quantum dark soliton; we call the graphs of the square amplitude and phase in the complex scalar field of a classical dark soliton versus the position coordinate the density profile and phase profile of classical dark soliton, respectively. 
Secondly, in time evolution the density profile of quantum dark soliton coincides with that of the corresponding classical dark soliton over the whole graph and they move together with the same velocity for a long period of time. 
Thirdly, for the matrix element of the bosonic field operator between two quantum dark soliton states where one state has $N-1$ particles and another $N$ particles, the profiles of the square amplitude and phase at the initial time graphically agree with those of classical dark soliton, respectively. In time evolution the profiles of square amplitude and phase overlap with those of classical dark soliton, respectively, over the whole region and the corresponding profiles move together in time for a long period of time. Here we remark that a classical dark soliton parametrized by elliptic functions becomes a standard dark soliton with open boundaries by simultaneously sending the system size to infinity and the elliptic modulus to unity. 
Furthermore,  in order to illustrate the method for constructing quantum dark solitons, in the 1D free fermions we show from the anti-commutation relations that a notch appears in the density profile of some superposition of one-hole excitations. Interestingly, the density profile of the fermionic state  coincides with that of quantum dark soliton for the 1D Bose gas in the large coupling case, i.e. near the TG limit,  not only at an initial time but also during the quantum dynamics for some period of time.

The time evolution of the expectation value of the local density operator in the 1D Bose gas should be important also from the renewed interest on the fundamentals of quantum statistical mechanics \cite{Tasaki98, Lebowitz, Popescu, Sugita, Reimann07}. The density profile of a quantum dark soliton state has initially a localized notch but collapses slowly to a flat profile in time evolution \cite{SKKD1}. The relaxation behavior is consistent with the viewpoints of equilibration of an isolated quantum system \cite{Rigol-GGE,Rigol-generic} and thermalization due to the typicality of quantum states \cite{TasakiNJP15}. 

We now argue that the density profile of quantum dark soliton has a finite lifetime but it is much longer than the quantum speed limit time of the quantum state in the weak coupling case. Here we remark that the lifetime of a generic state is given by the quantum speed limit time for it \cite{MT,ML,GLM2}. By observing the exact time evolution of the density profile of  quantum dark soliton we estimate its lifetime. We shall show in section 5  that the observed life time is inversely proportional to the coupling constant. We thus suggest that the localized density profile of a quantum dark soliton is much more stable than the density profile of a generic quantum state in the weak coupling case.  Here we remark that Girardeau and Wright discussed a permanent quantum soliton for impenetrable bosons in 1D \cite{Girardeau-Wright}, which corresponds to the infinite coupling case of the LL model. 

We also argue that the behavior of the wavefunctions is nontrivial in the weak coupling limit for the LL model. The wavefunctions do not simply become close to  
such wavefunctions  of the 1D free bosons that could be consistent with the mean-field picture, when the coupling constant approaches zero but it takes a nonzero value.  The exact Bethe ansatz wavefunctions consist of a large number of terms such as $N!$. Furthermore, we can show that many-body correlations increase if we increase the particle number $N$ while we keep the coupling constant small but fixed and the particle density constant. For instance, the zero mode fraction in the ground state of the 1D Bose gas, which we regard as the condensate fraction of BEC, becomes small and decreases to zero if the number of particles increases while the coupling constant and the particle density are  fixed \cite{BEC-1DBoseGas}. We suggest that it is also the case for quantum dark solitons i.e., that  some properties of  quantum dark solitons  
corresponding to classical dark solitons do not necessarily hold if we increase the particle number $N$ while we keep the coupling constant small but fixed and the density of particles fixed. In fact, it is not {\it a priori} clear how valid  the mean-field approximation is for 
quantum dark solitons even in the weak coupling case.  Here we remark that the breakdown of mean-field theory is addressed for impenetrable bosons \cite{Girardeau-Wright-breakdown}, and a long-wavelength theory beyond the mean-field approximation is discussed for low-dimensional bose liquids \cite{Kolomeisky}.   

Let us consider the Hamiltonian of the LL model for $N$ bosons with coupling constant $c$  
\begin{align}
{\cal H}_{\rm LL} 
= - \sum_{j=1}^{N} {\frac {\partial^2} {\partial x_j^2}}
+ 2c \sum_{j < k}^{N} \delta(x_j-x_k) . 
\label{h1}
\end{align}
Here we impose periodic boundary conditions with length $L$. 
We employ a system of units with $2m=\hbar =1$, where $m$ is the particle mass. 
We introduce the canonical Bose field operator $\hat{ \psi}(x,t)$ with the commutation relations \cite{Korepin}
\begin{align}
&[\hat{ \psi}(x,t), \hat{ \psi}^{\dagger}(x',t)]=\delta(x-x'), \\
&[\hat{ \psi}(x,t), \hat{ \psi}(x',t)]=[\hat{ \psi}^{\dagger}(x,t), \hat{ \psi}^{\dagger}(x',t)]=0. 
\end{align}
The second-quantized Hamiltonian of the LL model is written in terms of the field operator $\hat{ \psi}(x,t)$ as
\begin{align}
{\cal H}
= \int_{0}^{L} dx [ \partial_x \hat{ \psi}^{\dagger} \partial_x \hat{ \psi} + 
c \hat{ \psi}^{\dagger} \hat{ \psi}^{\dagger} \hat{ \psi} \hat{ \psi} - 
\mu \hat{ \psi}^{\dagger} \hat{ \psi} ],  
\label{h2}
\end{align}
where $\mu$ is the chemical potential. 
The Heisenberg equation of motion of this system has the form
\begin{align}
\i \partial_t \hat{ \psi} =  - \partial^2_x \hat{ \psi} 
+ 2c \hat{ \psi}^{\dagger} \hat{ \psi} \hat{ \psi} 
- \mu \hat{\psi}  .   
\label{eq:QNSE} 
\end{align}

In the classical limit where the quantum field operator $\hat{ \psi}(x,t)$ is replaced by 
a complex $c$-number field $\psi_C(x,t)$, equation (\ref{eq:QNSE}) 
becomes the following partial differential equation
\begin{align}
\i \partial_t \psi_C=- \partial^2_x \psi_{C} + 2c|\psi_C|^2\psi_C -\mu\psi_C . \label{eq:NSE}  
\end{align}
We call it the nonlinear Schr{\"o}dinger equation. The equation (\ref{eq:NSE}) can be solved by the inverse scattering method and has soliton solutions \cite{soliton} (see also \cite{Kawata}). We recall that it has dark soliton solutions for the repulsive interactions $c>0$, while it has bright soliton solutions for the attractive interactions $c<0$. For the attractive case, bright solitons are analytically derived from the quantum wave packets constructed through the Fourier transform of the bound states \cite{Wadati}. The construction was extended for optical fibers \cite{Refc}. Time evolution of quantum states of bright solitons was investigated theoretically \cite{Refb} and experimentally \cite{Refd}. The effect of quantum noises on the quantum states of bright solitons was also studied  \cite{Refe, Reff, Refg}.

The contents of the paper consists of the following. In section 2 we briefly introduce the notation for the Bethe ansatz, characterize the type II excitation branch, and then construct quantum dark soliton states. We argue that the structure of quantum dark soliton states is analogous to the Dirichlet kernel, and show for free fermions that the density profile of a superposition of one-hole excited states has a notch. It illustrates the construction of quantum dark soliton states. 
In section 3 we derive one-soliton solutions of the nonlinear Schr\"{o}dinger equation (\ref{eq:NSE}) under the  periodic boundary conditions. 
We express the classical dark solitons in terms of elliptic functions similarly as in Ref. \cite{Refa, Rina}. By sending the elliptic modulus $k$ to 1, the classical dark solitons 
under the periodic boundary conditions become the standard 
classical dark solitons of the nonlinear Schr{\"o}dinger equation with open boundaries. Furthermore, we calculate the logarithmic corrections to them, in particular, to the chemical potential and the soliton velocity, by which we can confirm the numerical estimates of the soliton parameters shown in section 4 at least approximately. 
In section 4 we show several corresponding aspects between the quantum dark soliton states and the corresponding classical dark solitons, which hold in the weak coupling case and should be rigorously valid when the coupling constant approaches zero with the density and the particle number $N$ fixed. Firstly, we show that the density profile of quantum dark soliton at an initial time is consistent with that of classical dark soliton. Secondly, we show that in exact time evolution the density profile of quantum dark soliton as a whole coincides with that of classical dark soliton for a long period of time. 
Thirdly, we show that both the square amplitude and phase profiles for the matrix element of the quantum field operator $\hat{\psi}(x, t)$ in the LL model between the quantum dark soliton states with $N-1$ and $N$ particles overlap with those of classical dark soliton, respectively, all over   the profiles. The square amplitude and phase profiles move in time evolution with exactly the same speed as those of classical dark soliton, respectively. 
Furthermore, we remark that the density profile of a sum of one-hole excitations in the 1D free fermions overlap  with that of quantum dark soliton in the 1D Bose gas with the large coupling constant, i.e. in a  regime close to the TG limit.  For the matrix element of the fermionic field operator between two such fermionic states  the square amplitude and phase profiles have several features in common with those of quantum dark soliton in the 1D Bose gas for the large coupling case, respectively, although they do not overlap each other.  We thus obtain useful tools to describe approximately the profiles of quantum dark soliton in the large coupling case. However, it is still quite nontrivial that the profiles of quantum dark soliton in the weak coupling case are consistent with those of classical dark soliton. 
In section 5 we argue that the lifetime of a notch in the density profile of quantum dark soliton is much longer than that of a generic state when the coupling constant $c$ is very small. We compare the lifetime of a notch in the density profile of quantum dark soliton with the quantum speed limit time, and show that the ratio of the former to the latter increases as the coupling constant decreases. It becomes infinitely large as the coupling constant approaches zero. Thus, although the localized density profile collapses in time evolution, we call the quantum states quantum dark soliton states. 

%
%
\section{Excited states of the 1D Bose gas with delta-function potentials}
\subsection{Bethe ansatz eigenwavefunctions}	
In the framework of the Bethe ansatz method, the Bethe ansatz wavefunction which is expressed in terms of quasi-momenta $k_1, k_2, \ldots, k_N$  
\begin{align}
&\vp_{k_1,\cdots,k_N}(x_1,\cd,x_N)
=
\frac{c^{N/2}}{\sqrt{N!}}
\(\prod_{j>\ell}^N\frac1{k_j-k_\ell}\)
\sum_{\sg\in S_N}^{N!}
A_\sg
\exp\[\i\sum_{j=1}^N k_{\sg_j} x_j\], \nn\db
&A_\sg=(-1)^\sg
\prod_{j>\ell}^N\[k_{\sg_j}-k_{\sg_\ell}-\i c \, \sign(x_j-x_\ell)\] 
\end{align}
gives an eigenstate of the LL model \eqref{h1}, 
if the quasi-momenta satisfy the Bethe ansatz equations 
\begin{align}
e^{\i k_j L} = \prod_{\ell \ne j}^{N} 
\frac
{k_j - k_{\ell} + \i c }
{k_j - k_{\ell} - \i c }
\quad
\text{for}
\quad
j=1, 2, \ldots, N.
\label{BAE1} 
\end{align}
In the logarithmic form, we have 
\begin{align}
k_j L = 2 \pi I_j - 2 \sum_{\ell \ne j}^{N} 
\arctan \left({\frac {k_j - k_{\ell}} c } \right)
\quad
\text{for}
\quad
j=1, 2, \ldots, N . 
\label{BAE} 
\end{align}
Here $I_j$'s are integers for odd $N$ and half-odd integers for even $N$. 
We call them the Bethe quantum numbers. 
The total momentum $P$ and energy eigenvalue $E$ are written 
in terms of the quasi-momenta as 
\begin{align}
P=\sm{j}{N}k_j=\frac {2 \pi} L \sum_{j=1}^{N} I_j, \quad E=\sm{j}{N}k_j^2. 
\end{align}
In the case of repulsive interaction $c>0$, 
the Bethe equations \eqref{BAE} have a unique real solution $k_1 < \cdots < k_N$ 
if we specify a set of Bethe quantum numbers $I_1<\cdots<I_N$ \cite{Korepin}. 

\begin{figure}
\includegraphics[width=0.9\columnwidth]{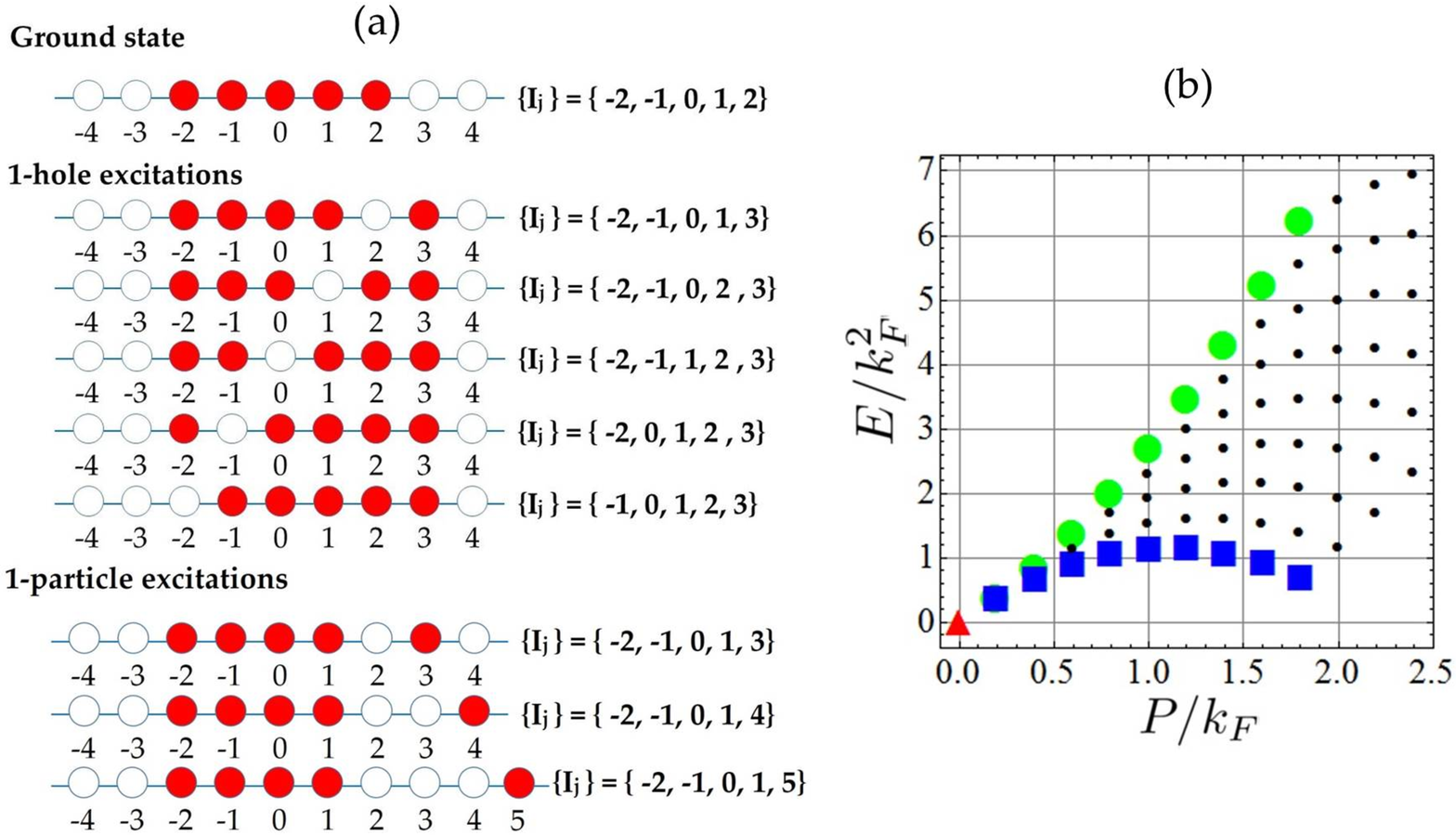}
\caption{ 
(a) Configurations of Bethe quantum numbers 
for the ground state, 1-hole excitations and 1-particle excitations 
in the case of $N=5$. 
(b) Dispersion relations of the 1-hole excitations and 1-particle excitations 
in the case of $N=5$. 
A triangle represents the ground state, 
big circles represent 1-particle excitations, 
and squares represent 1-hole excitations. 
Small dots are the 1-particle 1-hole excitations. 
}
\label{Ij1h1p}
\end{figure}

The configuration of the Bethe quantum numbers for the ground state is given by the regular array around the center: 
\begin{equation} 
I_j=-(N+1)/2 + j \qquad \mbox{for} \quad j =1, 2, \ldots, N. 
\end{equation}  
It is depicted in the top panel of Figure \ref{Ij1h1p} (a) in the case of $N=5$.  

We can construct low-lying excited states systematically by putting holes and placing particles in the regular array of the Bethe quantum numbers for the case of the ground state. The arrays of the Bethe quantum numbers for one-hole excitations are shown in the second panel of Figure \ref{Ij1h1p} (a). In each sequence there exists an unoccupied number in the middle of the sequence of the five red circles, which we call the one-hole. We remark that the one-hole excitations are also called the type II excitations \cite{Lieb-Liniger}. Similarly as in the second panel, the arrays of the Bethe quantum numbers for one-particle excitations are shown in the third panel of Figure \ref{Ij1h1p} (a). 
The right most red particle corresponds to the one-particle in each array. The one-particle excitations are also called the type I excitations \cite{Lieb-Liniger}.

The dispersion relation of the one-particle excitations and that of the one-hole excitations are plotted with filled circles and filled squares, respectively, in Figure \ref{Ij1h1p} (b). The former branch corresponds to the type I excitations and the latter branch the type II excitations. 

\subsection{Construction of quantum dark soliton states through type II excitations}
We now explain how we construct a series of quantum states \cite{SKKD1} through the type II excitations. 
Furthermore, we call them quantum dark soliton states. 

In the type II branch, for each integer $p$ in the set $\{0, 1, \ldots, N-1\}$, 
we consider momentum $P=2 \pi p/L$. 
We denote the normalized Bethe eigenstate of $N$ particles 
with total momentum $P$ by $|P, N\ket$, which we call a one-hole excited state. 
The Bethe quantum numbers $I_j$'s of the one-hole excitation $|P, N \ket$ are given by
\begin{align} 
I_j & = -(N+1)/2+j \quad \mbox{for} \quad 1 \leq j \leq N-p \nonumber \\ 
 & = -(N+1)/2+j+1 \quad \mbox{for} \quad N-p+1\leq j \leq N . 
\label{eq:Ij-one-hole}
\end{align} 
For any value of $X$ satisfying $0 \le X < L$ we define the coordinate state 
$|X, N \ket$ by the discrete Fourier transformation: 
\begin{align}
| X, N \ket := \frac 1 {\sqrt{N}} \sum_{p=0}^{N-1} \exp(- 2 \pi \i pX/L) \, | P, N \ket \, .  \label{eq:XN}
\end{align}
The density profile of this state $\bra X, N|\hpd(x)\hp(x)|X, N\ket$ shows a density notch as will be shown in section 4.

\subsection{Quantum state analogue of a truncated series of the delta function}

The formal structure of  quantum dark soliton states \eqref{eq:XN}
is analogous to a truncated series of the delta function.  
For periodic functions with period $L$ the delta function is expressed 
in the form of an infnite series   
\begin{equation}
\delta(x) = {\frac 1 {L}} 
\sum_{n=-\infty}^{\infty} \exp(2 \pi i n x/L )  \, . 
\end{equation} 
We now truncate it by keeping only such terms with 
integers $n$ satisfying $|n| \le N_c$ 
\begin{eqnarray}
D_{N_c}(x) & = & {\frac 1 {L} } 
\sum_{n=-N_c}^{N_c} \exp(2 \pi i n x/L ) \nonumber \\ 
& = & {\frac 1 {L}} \frac 
{\sin\left( (2N_c+1)\pi x/L \right)} {\sin \left(\pi x/L \right)} \, .  \label{eq:truncatedsum}
\end{eqnarray} 
It is called the Dirichlet kernel in the Fourier series analysis.  It has the peak around at $x=0$ with peak height $(2N_c+1)/L$ and oscillating behavior in the region for $|x| > L/(2N_c+1)$ with 
period $L/(2N_c+1)$. Here we assume that $N_c$ is independent of $L$. For $N_c \gg L$, the oscillations in the graph of the Dirichlet kernel become of very short wavelength, 
and we may call them ripples.

The integral of the squared amplitude of the Dirichlet kernel  
over the period $L$ is given by 
\begin{equation} 
\int_{-L/2}^{L/2} |D_{N_c}(x)|^2 dx  = \frac {2N_c+1} L \, .   
\end{equation}
We therefore normalize the function $D_{N_c}(x)$ 
by dividing it by the square root $\sqrt{(2N_c+1)/L}$  
\begin{equation}
\widetilde{D}_{N_c}(x) =  D_{N_c} (x)   \sqrt{ \frac L  {2N_c+1}} \, .   
\end{equation}
In terms of the normalized momentum eigenfunctions 
$\langle x | P \rangle = \exp(2 \pi i  px/L)/\sqrt{ L} $ where $P=2 \pi p/L$
the normalized function $\tilde{D}_{N_c}(x)$ is expressed as 
\begin{equation}
\widetilde{D}_{N_c}(x) =   {\frac 1  {\sqrt{2N_c+1}} } \sum_{p=-N_c}^{N_c} 
\langle x | P \rangle \, . \label{eq:xP}
\end{equation} 
It is analogous to the structure of the states of quantum dark solitons  \eqref{eq:XN},  
where  the number $2N_c+1$ in (\ref{eq:xP}) 
corresponds to the number of states $N$ in \eqref{eq:XN}, and 
the single-particle momentum eigenstates $|P \rangle$ to 
the one-hole excitations  $|P, N \rangle$ consisting of $N$ particles.  

By considering the phase factors $\exp(- 2 \pi i p X /L)$
in \eqref{eq:XN} the graphs are shifted by $X$ in the $x$-direction. 
We have $D_{N_c}(x-X)$ for the truncated series of the delta function 
modified with the phase factors.

\subsection{Density notch in the sum over one-hole excitations for free fermions}
 
Through the anti-commutation relation of the field operators  we now show
that a notch appears  in  the density profile of 
a linear combination of one-hole excited states in the 1D free fermions.

Under the  periodic conditions of length $L$  the  momenta are given by 
$k_{\alpha} = 2\pi n_{\alpha} /L$ with some integers $n_{\alpha}$ 
in the 1D free fermions.  
 We denote by $a_{\alpha}$ and $a_{\alpha}^{\dagger}$ for some integers $\alpha$ 
the creation and annihilation operators of the one-dimensional free fermions with momenta 
$k_{\alpha}=2 \pi n_{\alpha}/L$, respectively. We assume that an infinite series of integers $n_{\alpha}$ for $\alpha=1, 2, \ldots, $ cover all integers.   
The field operator and its conjugate are given by 
\begin{equation} 
\psi (x) = {\frac 1 {\sqrt{L}}} \sum_{\alpha=1}^{\infty} e^{ik_{\alpha} x } a_{\alpha} \, , \quad 
\psi^{\dagger} (x) = {\frac 1 {\sqrt{L}}} \sum_{\alpha=1}^{\infty} e^{-i k_{\alpha} x } a^{\dagger}_{\alpha} \, . \label{eq:field}
\end{equation}
Let us take a set of arbitrary integers $n_1, n_2, \ldots, n_M$ and consider the state 
of $M$ fermions with momenta $k_{\alpha}$'s  
\begin{equation}  
| M \rangle = \prod_{\ell=1}^{M} a_{\ell}^{\dagger} |0 \rangle .  \label{eq:stateM} 
\end{equation}
For simplicity, we may assume that $| M\rangle$ is the ground state.   
By applying the field operator $\psi(0)$ to it,  
we have  
\begin{eqnarray}
\psi(0) | M \rangle & = & \left( \frac 1 {\sqrt{L}} \sum_{\beta=1}^{\infty} a_{\beta} \right) 
\prod_{\ell=1}^{M} a_{\ell}^{\dagger} \,  |0 \rangle \nonumber \\ 
& & = \frac 1 {\sqrt{L}} \sum_{\beta=1}^{M}  (-1)^{\beta-1}  \prod_{\ell=1; \ell \ne \beta}^{M} 
a_{\ell}^{\dagger} \,  |0 \rangle \, . 
\end{eqnarray} 
Here we have assumed the ordering of fermionic operators as 
$\prod_{\ell=1}^{M} a_{\ell}^{\dagger} = a_1^{\dagger} a_2^{\dagger} \cdots a_M^{\dagger}$. 
We regard  states  $\prod_{\ell=1; \ell \ne \beta}^{M} 
a_{\ell}^{\dagger} \,  |0 \rangle$ as  `one-hole excitations' of $|M \rangle$. 
We define the state $|\Phi \rangle$ by the sum over all one-hole excitations 
derived from $| M \rangle$  
\begin{equation}
|\Phi \rangle =  {\frac 1 {\sqrt{M}}} \sum_{\beta=1}^{M}  (-1)^{\beta-1}  \prod_{\ell=1; \ell \ne \beta}^{M} 
a_{\ell}^{\dagger} \,  |0 \rangle \, .  \label{eq:Phi}
\end{equation}

Let us define the local density $\rho_{\Phi} (x)$ for the state $| \Phi \rangle$ of the 1D free fermions by 
\begin{equation} 
\rho_{\Phi}(x) = \langle \Phi | \psi^{\dagger}(x) \psi(x) | \Phi \rangle . 
\end{equation}
We now show that the local density vanishes at the origin: $\rho_{\Phi}(0)=0$ for the state. We first recall that the square of the field operator at the origin vanishes,  
$\psi(0)^2=0$, due to the anti-commutation relations of the field operators.  
Thus, we have 
\begin{eqnarray} 
\rho_{\Phi}(0) & = & \langle \Phi | \psi^{\dagger}(0)   \cdot \psi(0) | \Phi \rangle 
\nonumber \\  
& = & \langle \Phi | \psi^{\dagger}(0)  \cdot  \psi(0) \psi(0) | M \rangle   \sqrt{\frac M L }
\nonumber \\ 
& = & 0 \, . 
\end{eqnarray}
We can show that the integral of   the local density over the entire interval $L$ 
is given by a large positive value such as $M-1$   
\begin{equation}
\int_{-L/2}^{L/2} \rho_{\Phi}(x) dx = M-1 \, . 
\end{equation}
It follows that the local density cannot be always equal to zero, although it vanishes at the origin. 
Therefore, for the state $|\Phi \rangle$  which is given by  the sum over all the one-hole excited states of $| M \rangle$, the local density $\rho_{\Phi}(x)$ has a notch at least at the origin. 

In section 4 
we shall construct the free-fermionic analog of quantum dark soliton states for the 1D free fermions and show that the density profile coincides with those of the LL model if the coupling constant is large enough such as $c=100$.

%
%
\section{Classical dark solitons in the ring}
\subsection{Traveling-wave solutions expressed in terms of elliptic functions}
Let us assume the traveling wave solution with velocity $v$: 
$\psi_C(x,t)=\psi_C(x-vt)$ to the nonlinear Schr{\"o}dinger equation \eqref{eq:NSE}. Hereafter in section 3 we denote $\psi_C$ simply by $\psi$. 
We denote the time derivative by $\partial_t \psi=-v\psi'$ and the second spatial derivative by $\partial_{x}^{2} \psi=\psi''$, where $\psi'$ denotes $\d \psi /\d \theta$ with $\theta =x -vt$. We therefore have 
\begin{align}
\psi''-\i v\psi'+\mu\psi-2c|\psi|^2\psi=0. 
\label{eq2}
\end{align}
We denote the solution of eq. (\ref{eq2}) by $\psi(x)$ 
as a function of $x$, for simplicity. 
We express the complex scalar field $\psi(x)$ in terms of the amplitude 
$\sqrt{ \rho(x)}$ and the phase $\vp(x)$, both of which are real, as   
\begin{align}
\psi(x)=\sqrt{\rho(x)}e^{\i\vp(x)}. \label{eq:AP}
\end{align}
By substituting (\ref{eq:AP}) into equation \eqref{eq2} we derive 
a pair of coupled equations from the real and imaginary parts, respectively. 
\begin{align}
&\dfrac{(\sqrt{\rho})''}{\sqrt{\rho}}
-(\vp')^2+v\vp'+\mu-2c\rho
=0, 
\label{re}
\\
&2\vp'
\dfrac{(\sqrt{\rho})'}{\sqrt{\rho}}
+\vp''-v
\dfrac{(\sqrt{\rho})'}{\sqrt{\rho}}
=0. 
\label{im}
\end{align}
It follows from equation \eqref{im} that we have $(\rho\vp'-v\rho/2)'=0$. 
By integrating it once with respect to $x$, we have
\begin{align}
\vp'(x)=\frac{v}2+\frac{W}{\rho(x)} . 
\label{vpd}
\end{align}
Here $W$ denotes a constant of integration. 
Substituting (\ref{vpd}) into \eqref{re} and multiplying it by $\sqrt{\rho}(\sqrt{\rho})'$, we have
\begin{align}
(\sqrt{\rho})'(\sqrt{\rho})''
+\(\mu+\frac{v^2}4\)\sqrt{\rho}(\sqrt{\rho})'
-W^2(\sqrt{\rho})^{-3}(\sqrt{\rho})'
-2c(\sqrt{\rho})^{3}(\sqrt{\rho})'
=0. 
\end{align}
Integrating this equation with respect to $x$, we have 
\begin{align}
\frac12(\sqrt{\rho}')^2
+\(\frac{\mu}{2}+\frac{v^2}8\)(\sqrt{\rho})^2
+\frac{W^2}{2}(\sqrt{\rho})^{-2}
-\frac{c}{2}(\sqrt{\rho})^{4}
=V, 
\end{align}
where $V$ is another constant of integration. Thus, we have
\begin{equation}
\(\frac{\rho'}{2}\)^2+U(\rho)=0, \label{eq:U}
\end{equation}
where the potential function $U(\rho)$ is given by 
\begin{equation} 
U(\rho)=-c\rho^3+\(\mu+\frac{v^2}{4}\)\rho^2-2V\rho+W^2. 
\end{equation}
	
Let us assume that the cubic equation: $U(\rho)=0$ has three distinct real roots, $a_1$, $a_2$, and $a_3$. We put them in increasing order: $a_1 < a_2 < a_3$.  The coefficients of the cubic equation are expressed in terms of the roots as follows. 
\begin{align} 
&{\frac 1 c} \left(\mu + \frac {v^2} 4 \right)  =  a_1 + a_2 + a_3 \, , \nonumber \\ 
&\frac {2V} c  =  a_1 a_2 + a_2 a_3 + a_3 a_1 \, , \nonumber \\ 
&{\frac 1 c} W^2  =  a_1 a_2 a_3 \, . 
\label{eq:coefficients} 
\end{align} 

We now derive a solution $\rho(x)$ to equation (\ref{eq:U}) such that it satisfies $a_1 \le \rho(x) \le a_2$ in the interval of $x$ with $-L/2 \le x \le L/2$. We set the initial condition: 
$\rho(x=0)=a_1$, for simplicity. It thus follows from equation (\ref{eq:U}) that we have 
\begin{align}
\int_{a_1}^{\rho(x)}\frac{\d r}{\sqrt{-U(r)}}=2x  
 \label{eq:integral-x} \qquad \left( a_1 \leq \rho(x) \leq a_2 \right). 
\end{align}
Let us define the modulus of Jacobi's elliptic functions, $k$, by 
\begin{align}
k=\sqrt{\frac{a_2-a_1}{a_3-a_1}} . 
\label{r1}
\end{align}
It is clear that $0 < k < 1$.  
Through the transformation of variables from $r$ to $z$ 
\begin{align}
r=a_1+(a_2-a_1)z^2 \qquad (0 \leq z \leq 1) 
\end{align}
we express the integral (\ref{eq:integral-x}) in terms of the elliptic integral of the first kind: 
\begin{align}
2x =
\frac2{\sqrt{c}\sqrt{a_3-a_1}}
\int_{0}^{\sqrt{\frac{\rho-a_1}{a_2-a_1}}}
\frac{\d z}
{\sqrt{(1-z^2)(1-k^2z^2)}} \, . 
\label{eq:EllipticIntegral}
\end{align}
We therefore have the solution to eq. (\ref{eq:U}) in terms of Jacobi's elliptic function 
$\sn(u, k)$  
\begin{equation} 
\rho(x)=a_1+ (a_2-a_1)\sn^2\(\sqrt{c} \sqrt{a_3-a_1} x,k\) . \label{eq:rho}
\end{equation} 

We now integrate equation \eqref{vpd}. Here we remark that 
the constant $W$ is given by $W=\pm\sqrt{ca_1a_2a_3}$. 
In terms of the elliptic integral of the third kind \eqref{f3} we have 
\begin{align}
\vp(x)=\vp(0)+\frac{v}2x
\pm
\frac{\sqrt{a_2a_3}}{\sqrt{a_1}\sqrt{a_3-a_1}}
\Pi\(1-a_2/a_1,\am\(\sqrt{c} \sqrt{a_3-a_1} x\),k\), 
\end{align} 
where $\vp(0)$ corresponds to the constant of integration. 
\subsection{Classical dark solitons expressed in terms of elliptic integrals} 

We now express $a_1$, $a_2$ and $a_3$ in terms of the complete elliptic integrals. 
Here we consider the periodic boundary conditions (PBC) 
for the square amplitude $\rho(x)$ and the phase $\varphi(x)$ of the classical complex scalar field $\psi(x, t)$, and the normalization condition of the square amplitude $\rho(x)$. 
Then, the elliptic modulus $k$ is related to the depth of the classical dark soliton, 
i.e. the minimum value $a_1$ of the square amplitude $\rho(x)$. 
We express the soliton velocity $v$ in terms of the elliptic integral of the second kind, 
and determine the chemical potential $\mu$. 
\subsubsection{PBCs for the square amplitude and the phase of  classical dark soliton}
Suppose that the square amplitude $\rho(x)$ of 
a classical dark soliton increases from $a_1$ to $a_2$ and then 
returns back to $a_1$with period $L$. 
We therefore assign the conditions at the initial and the middle points as 
\begin{equation} 
\rho(x=0)=a_1  \, , \qquad \rho(x=L/2)=a_2 .
\label{eq:conditions}
\end{equation} 
It follows from \eqref{eq:EllipticIntegral} that 
in terms of the complete elliptic integral of the first kind, $K(k)$, we have
\begin{equation} 
\sqrt{c (a_3-a_1)} = \frac {2 K(k)} L \, . \label{r2} 
\end{equation} 
We have $a_3-a_1={4K^2}/{(c L^2)}$.
Here and hereafter we often denote $K(k)$ by $K$, 
suppressing the modulus $k$. 

We assume the periodic boundary condition for the phase: $\varphi(x=L) = \varphi(0)$. 
Suppose that the soliton velocity is positive: $v>0$. 
It follows that we have $W<0$ 
and the velocity $v$ is expressed in terms of the complete elliptic integral of the third kind as 
\begin{align}
v=
\frac{4\sqrt{a_2a_3}}{L\sqrt{a_1}\sqrt{a_3-a_1}}
\Pi\(1-a_2/a_1,k\). 
\end{align}
\subsubsection{Normalization condition for the square amplitude}
We impose the normalization condition to the square amplitude $\rho(x)$ 
\begin{align}
\frac1L\int^{L/2}_{-L/2} \rho(x)\d x= \nt , \label{eq:normal} 
\end{align}
where $\nt$ is the density of the number of particles $N$, i.e. we have $n=N/L$. 
By the formula \eqref{dndef} we transform equation (\ref{eq:rho}) 
in terms of Jacobi's $\dn$ function into the following. 
\begin{align}
\rho(x)=a_3-(a_3-a_1)\dn^2 \( \sqrt{c} \sqrt{a_3-a_1} \, x,k\). 
\label{eq:sq-amp-dn}
\end{align}
Thus, by the formula \eqref{f2} 
in terms of the complete elliptic integral of the second kind $E(k)$ we have 
\begin{align}
\nt=a_3-\frac{4KE}{L^2c}. 
\label{r3}
\end{align}
\subsubsection{Parametrization of $a_1$, $a_2$ and $a_3$ in terms of complete elliptic integrals}
By solving equations \eqref{r1}, \eqref{r2} and \eqref{r3} 
with respect to the roots $a_1$, $a_2$ and $a_3$, we express them in terms of the complete elliptic integrals as follows: 
\begin{align}
&a_1=\nt+\frac{4K(E-K)}{L^2c}, \label{eq:a1} \\
&a_2=\nt+\frac{4K(E-(1-k^2)K)}{L^2c}, \\
&a_3=\nt+\frac{4KE}{L^2c}. 
\end{align}
Since the minimum value of the amplitude must be non-negative, i.e. $a_1 \ge 0$, 
we assign the following condition
\begin{equation} 
\frac {4 K^2} {\nt c L^2} \le {\frac 1 {1-E/K}} \, . \label{eq:nonnegative}
\end{equation} 
Thus, if the coupling constant $c$, the system size $L$, the density $n$ 
and the elliptic modulus $k$ satisfy condition (\ref{eq:nonnegative}), 
the classical dark soliton exists as a solution of the nonlinear Schr{\"o}dinger 
equation \eqref{eq:NSE} (i.e. the GP equation). 
The soliton velocity $v$ and the chemical potential $\mu$ are given by 
\begin{align}
&v=
\frac{4\sqrt{a_2a_3}}{L\sqrt{a_1}\sqrt{a_3-a_1}}
\Pi\(1-a_2/a_1,k\), \label{eq:v} \\
&\mu=3\nt c-\frac{v^2}4+\frac{4K(3E-(2-k^2)K)}{L^2}. \label{eq:mu} 
\end{align}

\subsubsection{Numerical determination of the parameters of classical dark solitons}

Let us explain how we evaluate the parameters of a dark soliton such as 
$n$ and $k$ in order to compare the profiles of the dark soliton with those of  a given quantum dark soliton.  Here we remark that when a quantum state is given, parameters $c$, $L$ and the particle number $N$ have been specified. 
We determine the elliptic modulus $k$ from the depth of the dark soliton, i.e. $a_1$: 
We solve equation \eqref{eq:a1} as an equation for modulus $k$. 
We assign   the value $N/L$ on the parameter of density $n$ for the density profile of a quantum state.  However, we determine the density $n$ numerically 
for the squared amplitude profile of the matrix element of the bosonic field operator between different quantum states, since it may take a value smaller than $N/L$.

For the density profile of a quantum state $|\Phi \ket$, the integral of the expectation value $\bra \Phi| \hat{\rho}(x) |\Phi \ket$ of the local density operator 
over the whole space is equal to $N/L$.  
However, for the matrix element $\bra \Psi | {\hat \psi}(x) | \Phi \ket$ of the quantum field operator ${\hat \psi}(x) $ between two states $|\Phi \ket$ and $| \Psi \ket$, the integral of the squared amplitude $|\bra \Psi | {\hat \psi}(x) | \Phi \ket|^2$ over the whole space may be smaller than the value of $N/L$. 
In this case we determine the parameter $n$ numerically by taking the integral of the squared amplitude $| \bra \Psi | {\hat \psi}(x) | \Phi \ket|^2$ over the whole space.  
Here we remark that the symbols will be defined in section 4.1.1.  

In summary, after we fix the value of density $n$, we solve equation \eqref{eq:a1} as a function of modulus $k$ and estimate its numerical value very precisely. 
In section 4 we shall make the profiles of classical dark soliton  
for given density profiles of quantum dark soliton by the method explained in the above. 
\subsubsection{Critical velocity}
We define the function $f(k)$ by $f(k):=K(k) \[3E(k)-(2-k^2)K(k)\]$.  
It appears in the third term of the expression \eqref{eq:mu} for the chemical potential $\mu$. We can show that it is a monotonically decreasing function of modulus $k$, which yields the inequality $f(k) \leq f(0)=\pi^2/4$. This gives an upper bound for the absolute value of soliton velocity $v$: $|v| \leq v_c$, where 
\begin{align}
v_c=\sqrt{(2\pi/L)^2+4(3\nt c-\mu)} \, . 
\end{align}
Here we remark that the critical velocity is defined for the system of a finite size.
We shall denote by $v_{c, \infty}$ the critical velocity for an infinite system.

\subsection{Asymptotic behavior of a classical dark soliton} 
\subsubsection{Square amplitude $\rho(x)$ and the phase $\varphi(x)$ expressed in terms of elliptic functions} 

We now express a classical dark soliton explicitly in terms of the complete elliptic integrals. 
The square amplitude \eqref{eq:sq-amp-dn} is given by 
\begin{equation}
\rho(x)= n - {\frac {4 KE} {c L^2}} - {\frac {4 K^2} {c L^2}} \dn^2 \left( \frac {2K x} L , k \right) \, . 
\label{eq:rho-EK}
\end{equation} 
We shall show that the second term in the right-hand side of equation \eqref{eq:rho-EK} 
leads to the logarithmic correction associated with the conservation of the number of particles. 
Let us introduce a parameter $\beta_k$ by 
\begin{equation} 
\beta_k = \frac {4 K^2} {n c L^2} \, . 
\end{equation}
We denote the argument $2Kx/L$ in \eqref{eq:rho} by $u$. Then, the square amplitude is expressed as follows: 
\begin{equation} 
\rho(x) = n \left( 1 - \beta_k  {\frac E K} - \beta_k \, \dn^2 (u, k) \right) \, . 
\end{equation} 

We now show that the phase $\varphi(x)$ is expressed in terms of Jacobi's Theta function. 
Let us define a pure imaginary number $a= \i \alpha$ with $\alpha > 0$ by 
\begin{equation} 
\sqrt{1 - a_2/a_1} = k \, \sn(\i \alpha, k) \, . \label{eq:alpha} 
\end{equation} 
By formula (\ref{eq:EI3}) we evaluate the elliptic integral of the third kind with Jacobi's Zeta and Theta functions \cite{Copson}, and we express the phase $\varphi(x)$ as the logarithm of a ratio of Theta functions 
\begin{equation}
\varphi(x)-\varphi(0)= - {\frac \i 2} 
\log \frac {\Theta\({\frac {2K x} L} - \i \alpha\)} {\Theta\({\frac {2K x} L} + \i \alpha\)} \, .    
\label{eq:phase-theta}
\end{equation} 
\subsubsection{Parameters in the asymptotic expansion with respect to the system size $L$}
Let us consider the limit of sending the system size $L$ to infinity ($L \to \infty$)  
and the modulus $k$ to 1 ($k \to 1$) simultaneously 
so that the ratio $K(k)/L$ is kept constant. 
We denote the ratio by $b$: $b=K/L$. 
Here we remark that we keep the density $\nt$ constant. 
We define $\beta$ by the simultaneous limit of sending $L$ to infinity 
and $k$ to 1 with $K/L=b$. 
\begin{equation} 
\beta = \lim_{k \to 1, L \to \infty} \beta_{k} = \frac {4 b^2} {n c} \, . 
\end{equation}   

Let us introduce the complementary modulus $k^{'}$ by $k^{'}= \sqrt{1-k^2}$. 
We define $K^{'}(k)$ and $E^{'}(k)$ by $K^{'}(k)= K(k^{'})$ and $E^{'}(k)= E(k^{'})$, 
respectively. We remark that $K$ and $\i K^{'}$ give quarter periods of Jacobi's elliptic functions. 
We introduce a parameter $p$ by $p= \exp(-\pi K/K^{'})$, which is small when modulus $k$ is close to 1. 
Making use of formulae \eqref{eq:EandK}, \eqref{eq:KK'}, \eqref{eq:kk'} 
and Legendre's relation \eqref{eq:Legendre} we expand $K$'s and $E$'s with respect to $p$ as follows: 
\begin{align} 
&K =  \frac 1 2 \log(1/p) ( 1 + 4p + \cdots ) \, , \nn \\
&K^{'} = \frac {\pi} 2 ( 1+ 4p + \cdots ) \, , \nn \\
& E = 1+ 4p \log(1/p) - 4p + \cdots  \, , \nn \\ 
&E^{'} = \frac {\pi} 2 ( 1- 4p + \cdots ) \, , \nn \\ 
&k^2 = 1 - 16 p + \cdots . 
\label{eq:asymptotic}
\end{align} 
We have $p \simeq \exp(- 2b L)$, since we have fixed the ratio $K/L=b$.  
\subsubsection{Asymptotic expansion of the velocity and the chemical potential when $k$ approaches 1}
We recall that the soliton velocity $v$ is expressed in terms of the 
complete elliptic integral of the third kind \eqref{eq:v}. 
Through \eqref{eq:cei3} we express it in terms of Jacobi's Zeta function as follows: 
\begin{equation} 
 \frac v {2 \sqrt{n c}} = \sqrt{ \frac {a_2 a_3} {n a_1}} + \i \sqrt{\beta_k} \, Z(\i \alpha, k) \, . 
\label{eq:v-Zeta}
\end{equation}
Here we recall that parameter $\alpha$ has been defined by \eqref{eq:alpha}. 

Let us now evaluate $Z(\i \alpha, k)$. 
We first express the left-hand side of \eqref{eq:alpha} in terms of $\beta_k$ as
\begin{equation} 
\sqrt{1 - a_2/a_1} = \i \, k \, \sqrt{ {\beta_k}/ {(1 - \beta_k + \beta_k E/K )} }. 
\end{equation} 
When the modulus $k$ is close to 1, 
by making use of Jacobi's imaginary transformation of $\sn$ function: 
$\sn(\i \alpha, k) = \i \, {\sn( \alpha, k^{'})}/{\cn(\alpha, k^{'})}$ 
and then by expressing it in terms of Jacobi's Eta and Theta functions, we show 
\begin{equation} 
\tan \left( \frac {\pi \alpha} {2 K^{'}} \right) = k^{1/2} \sqrt{\frac {\beta} {1-\beta + \beta E/K} }
+ O(p^2) . \label{eq:tan-alpha}
\end{equation} 
Through \eqref{eq:Zeta-p} we express the value of Zeta function $Z(a, k)$ in terms of $a= \i \alpha$ as follows. 
\begin{equation}
Z(a, k) = \i \left( \frac {\pi} {2 K^{'}} 
\tan \left( \frac {\pi \alpha} {2 K^{'}} \right) 
- {\frac { \pi \alpha} {2 K K^{'} }} \right) + O(p^2) .  
\label{eq:Z-tan-alpha} 
\end{equation} 
Applying the expansions \eqref{eq:tan-alpha} and \eqref{eq:Z-tan-alpha} to \eqref{eq:v-Zeta}, 
we evaluate the velocity $v$ upto the order of logarithmic terms 
\begin{align}
\frac v {2 \sqrt{n c}} = \sqrt{1- \beta} + \sqrt{\beta} 
\left( \sqrt{ \frac {\beta} {1-\beta} } + 2 \arctan \left( \sqrt{\frac {\beta} {1-\beta} } \right) 
\right) \frac 1 {\log(1/p)} + \cdots  \, . 
\end{align} 
and through \eqref{eq:mu} we evaluate the chemical potential up to 
 the order of logarithmic terms 
\begin{align}
\frac {\mu} {2 n c} = 1 + 2 \sqrt{\beta(1- \beta)} 
\left( \sqrt{ \frac {\beta} {1-\beta} } - \arctan \left( \sqrt{\frac {\beta} {1-\beta} } \right) 
\right) \frac 1 {\log(1/p)} + \cdots \, . 
\end{align} 

It follows that chemical potential $\mu$ approaches $2 n c$ 
when we send system size $L$ to infinity. We denote the limiting value by $\mu_{\infty}$. 
Here we remark that the logarithmic corrections correspond to 
the finite size corrections since we have $\log(1/p) = 2 b L$. Moreover, 
the chemical potential $\mu$ is larger than $\mu_{\infty} = 2 n c$ 
when the system size is finite since we have $x > \arctan x $ for $x > 0$. 
We can confirm such behavior by checking the numerical estimates for the parameters of
classical dark solitons listed in section 4. 
\subsubsection{Asymptotic expansion of the square amplitude of the classical complex scalar field}
Through Jacobi's imaginary transformation for $\dn$ function we have 
\begin{equation} 
\dn(u, k) = \sech \left( \frac { \pi u} {2 K^{'}} \right)
\prod_{n=1}^{\infty} 
{\frac { 1 + 2 p^{2n-1} \cosh ( \pi u/K^{'}) + p^{4n-2} } 
      {1 + 2 p^{2n} \cosh( \pi u/K^{'} ) + p^{4n} } }
\prod_{n=1}^{\infty} \left( \frac {1+p^{2n}} {1+ p^{2n-1}} \right)^2 \, . 
\end{equation}
We therefore have 
\begin{equation} 
\rho(x)/n = 1 - \beta \sech^2 ( 2 b x ) - 2 \beta/\log(1/p) + \cdots \, . 
\end{equation} 
It is easy to show that the asymptotic expansion of the square amplitude 
up to the order of $1/\log p$, i.e. the order of $1/L$, satisfies the normalization condition \eqref{eq:normal}. 

At $k=1$, the square amplitude $\rho(x)$ is expressed in terms of 
 $\b=4b^2/(\nt c)$ by 
\begin{align}
\rho(x)=\nt(1- \b \, \sech^2 (2bx) ) \, . 
\end{align}
Thus, the critical velocity $v_c$ with the infinite system size ($L = \infty$) 
is given by $v_{c, \infty}=2\sqrt{3\nt c-\mu_{\infty}}= 2 \sqrt{\nt c}$. 
Here we recall $\mu_{\infty} = 2 nc$. 
It follows that we have
\begin{align}
&\rho(x)=\(\frac{\mu_{\infty}}{2c}\) \left\{1-\b\sech^2 \[\( \frac{\b\mu_{\infty}}2 \)^{1/2} x\] \right\} .  
\end{align}
Here $\beta$ is expressed as $\b=1-(v/v_{c, \infty})^2$. 
We have thus derived the square amplitude of one-soliton solution due to Tsuzuki \cite{soliton}. 
\subsubsection{Asymptotic expansion of the phase of the classical complex scalar field}
Applying \eqref{eq:Theta-p} to \eqref{eq:phase-theta} we derive the following expansion 
in terms of $p= \exp(-2bL)$ for any system size $L$: 
\begin{align} 
\varphi(x) - \varphi(0) & =  \frac {\pi \alpha u} {2 K K ^{'}} 
+ {\frac 1 {2 \i}} \log \left( \frac {1 - \i \tan (\pi \alpha/2 K^{'}) \tanh (\pi u/2K^{'}) } 
                       {1 + \i \tan (\pi \alpha/2 K^{'}) \tanh (\pi u/2K^{'}) } \right) \nonumber \\ 
&  + \sum_{n=1}^{\infty} 
\frac 1 {2 \i} \log \left( 
\frac {1 + 2 p^{2n} \cosh \left( \pi (u - \i \alpha)/K^{'} \right) + p^{4n} }
    {1 + 2 p^{2n} \cosh \left( \pi (u + \i \alpha)/K^{'} \right) + p^{4n} }
\right) \, . 
\end{align} 
Making use of \eqref{eq:tan-alpha} we derive the expansion of $\varphi(x)$ 
up to the order of the inverse of $\log(1/p)$ as follows: 
\begin{align} 
\varphi(x) - \varphi(0) & =  
\frac 1{2 \i} \log \left( \frac {\sqrt{1-\beta} - \i \sqrt{\beta} \tanh(2b x)} 
{\sqrt{1-\beta} + \i \sqrt{\beta} \tanh(2b x)} 
 \right) \nn \\ 
& + \left\{  4 b x \, \arctan \left( \sqrt{ \frac {\beta} {1 - \beta}} \right) 
+ {\frac { \beta \sqrt{\beta/(1-\beta)} \tanh^2 2bx }
     {1 - \beta \sech^2( 2bx)}} \right\} /\log(1/p) + \cdots  . 
\end{align} 
Therefore, in the limit of sending $k$ to 1, 
we obtain the phase of the dark soliton solution by Tsuzuki \cite{soliton}: 
\begin{equation}
e^{ \i (\varphi(x)- \varphi(0) )} = 
\frac {\sqrt{1-\beta} - \i \sqrt{\beta} \tanh 2bx }
{\sqrt{1 - \beta \sech^2 2bx }} \, .  
\end{equation} 
Thus, we have shown that the classical dark soliton under the PBCs approaches the dark soliton solution by Tsuzuki \cite{soliton} through the simultaneous limit of sending the system size to infinity and the modulus $k$ to 1. 

%
%
\section{Aspects of quantum states corresponding to classical solitons}
\subsection{Density profile of quantum and classical dark solitons}
\subsubsection{Formula of form factors of the local density operator}
We define the local density operator $\hat{\rho}(x)$ by $\hat{\rho}(x):=\hpd(x) \hp(x)$, 
in the second-quantized system of the 1D Bose gas interacting through the delta-function potentials \eqref{h2}. 
Here we denote by $\hp(x)$ the field operator at the initial time $t=0$: $\hp(x, t=0)$.  
We now consider the graph of the expectation value of   
the local density operator for a quantum state $| \Phi \ket$ 
(i.e. $\bra \Phi| \hat{\rho}(x)| \Phi \ket$) versus the position coordinate $x$. 
Here we recall that we call it the density profile of the state $| \Phi \ket$. 

We evaluate the expectation value of the local density operator 
for the state $|X,N \rangle$ by the form factor expansion. 
We express the expectation value as the sum over the form factors between the Bethe eigenstates  
\begin{align}
\bra X, N|\hat{\rho}(x)|X, N\ket
&=
\frac{1}{N}\sum^{N-1}_{p,p'=0}
\exp\[2 \pi \i (p-p') \frac{(x-X)}{L} \]
\bra P', N|\hat{\rho}(0)|P, N\ket \, .  \label{eq:DP-formfactor}
\end{align}
Here $|P, N\ket$ and $|P', N\ket$ are the normalized Bethe eigenstates of $N$ particles 
in the type II branch (i.e. they are one-hole excitations)
and have total momentum $P=2 \pi p/L$ and $P'=2 \pi p'/L$, respectively. 
Here the form factors $\bra P', N|\hat{\rho}(0)|P, N\ket$
are effectively calculated \cite{SKKD1} by the determinant formula for the norms of Bethe eigenstates \cite{GK} 
and that of the form factors of the density operator \cite{Slavnov, Caux2007} 
\begin{align}
\bra P',N|\hat{\rho}(0)|P,N\ket
=(-1)^{N(N+1)/2}(P-P')
\(\prod^N_{j,\ell=1}\frac{1}{k'_j-k_\ell}\) 
\( \prod^N_{j>\ell}k_{j,\ell}k'_{j,\ell}\sqrt{\frac{{\hat K}(k'_{j,\ell})}{{\hat K}(k_{j,\ell})} } \)
\frac{\det U(k,k')}{\sqrt{\det G(k)\det G(k')}}, 
\label{eq:Slavnov_density}
\end{align}
where the quasimomenta $\{k_1,\cdots,k_N\}$ and $\{k'_1,\cdots,k'_N\}$ 
lead to the eigenstates $|P\ket$ and $|P'\ket$, respectively. 
We use the abbreviations $k_{j,\ell}:=k_j-k_\ell$ and $k'_{j,\ell}:=k'_j-k'_\ell$. 
The kernel ${\hat K}(k)$ is defined by ${\hat K}(k)=2c/(k^2+c^2)$. 
The matrix $G(k)$ is called the Gaudin matrix, whose $(j,\ell)$ th element is given by 
\begin{equation} 
G(k)_{j,\ell}=\delta_{j,\ell}\[L+\sum_{m=1}^N {\hat K}(k_{j,m})\]-{\hat K}(k_{j,\ell})
\quad \mbox{for} \quad j, \ell=1, 2, \cdots, N . \label{eq:Gaudin}
\end{equation} 
The matrix elements of the $(N-1)$ by $(N-1)$ matrix $U(k,k')$ are given by 
\begin{align}
U(k,k')_{j,\ell}=2\delta_{j\ell}\text{Im}\[\prod^N_{a=1}
\frac{k'_a-k_j + \i c}{k_a-k_j + \i c}\]
+\frac{\prod^N_{a=1}(k'_a-k_j)}{\prod^N_{a\neq j}(k_a-k_j)} 
\({\hat K}(k_{j,\ell})-{\hat K}(k_{N,\ell})\). 
\label{eq:matrixU_density}
\end{align}

\subsubsection{Density profiles of quantum and classical dark solitons at the initial  time}


\begin{figure}
\includegraphics[width=0.9\columnwidth]{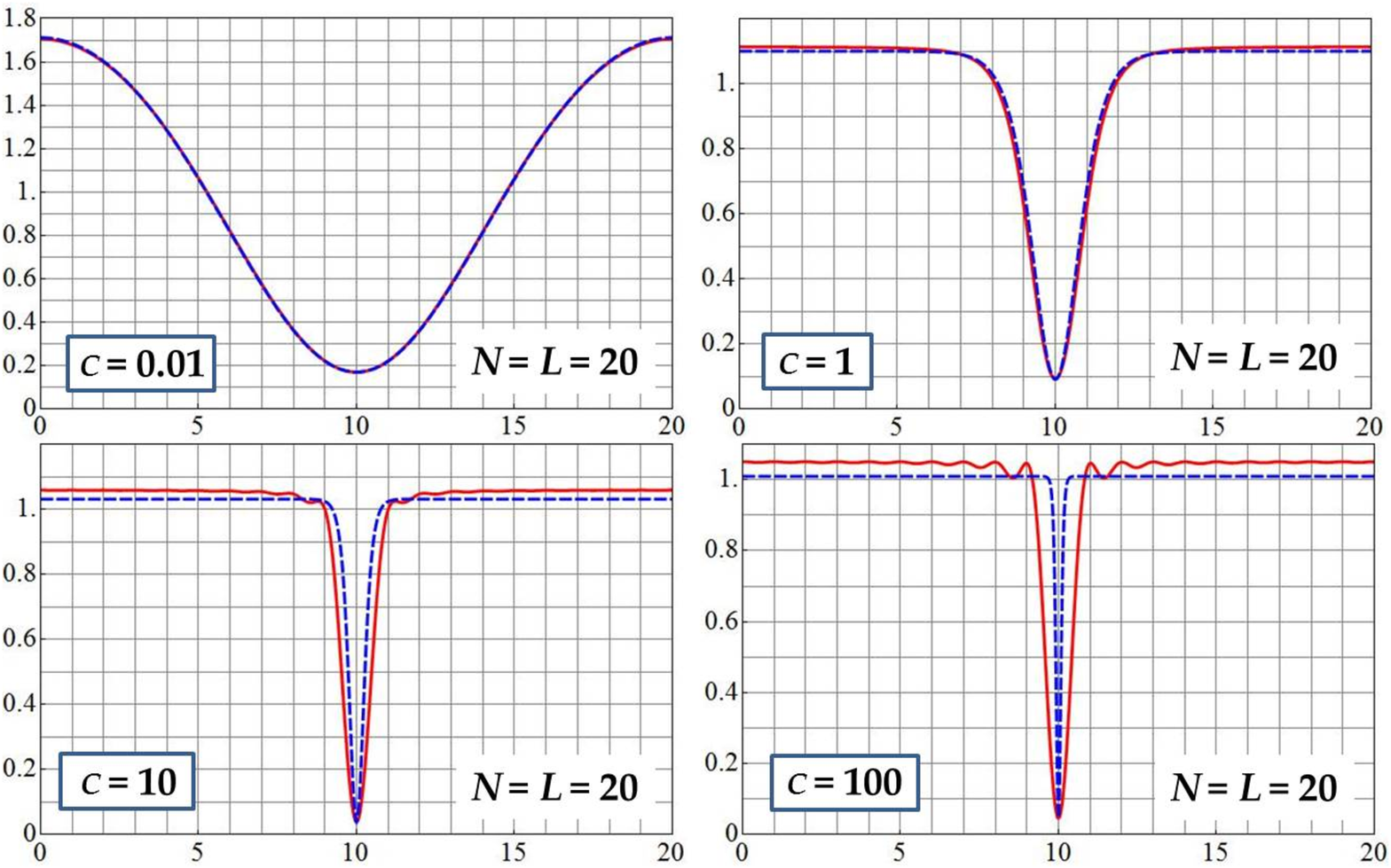}
\caption{
Density profile of quantum dark soliton with $N=L=20$, i.e. $\bra X, N|\hat{\rho}(x)| X, N\ket$ versus $x$, is shown by a red solid line to each value 
of $c$ with $c=0.01, 1, 10$ and 100.  
The density profile of 
classical dark soliton 
$|\mfperi(x)|^2 = \rho(x-L/2)$ 
with parameters shown in Table \ref{density_table20-T1} is plotted 
with a blue broken line to each value of $c$ with $c=0.01, 1, 10$ and 100. }
\label{density20}
\end{figure}

\begin{figure}
\includegraphics[width=0.9\columnwidth]{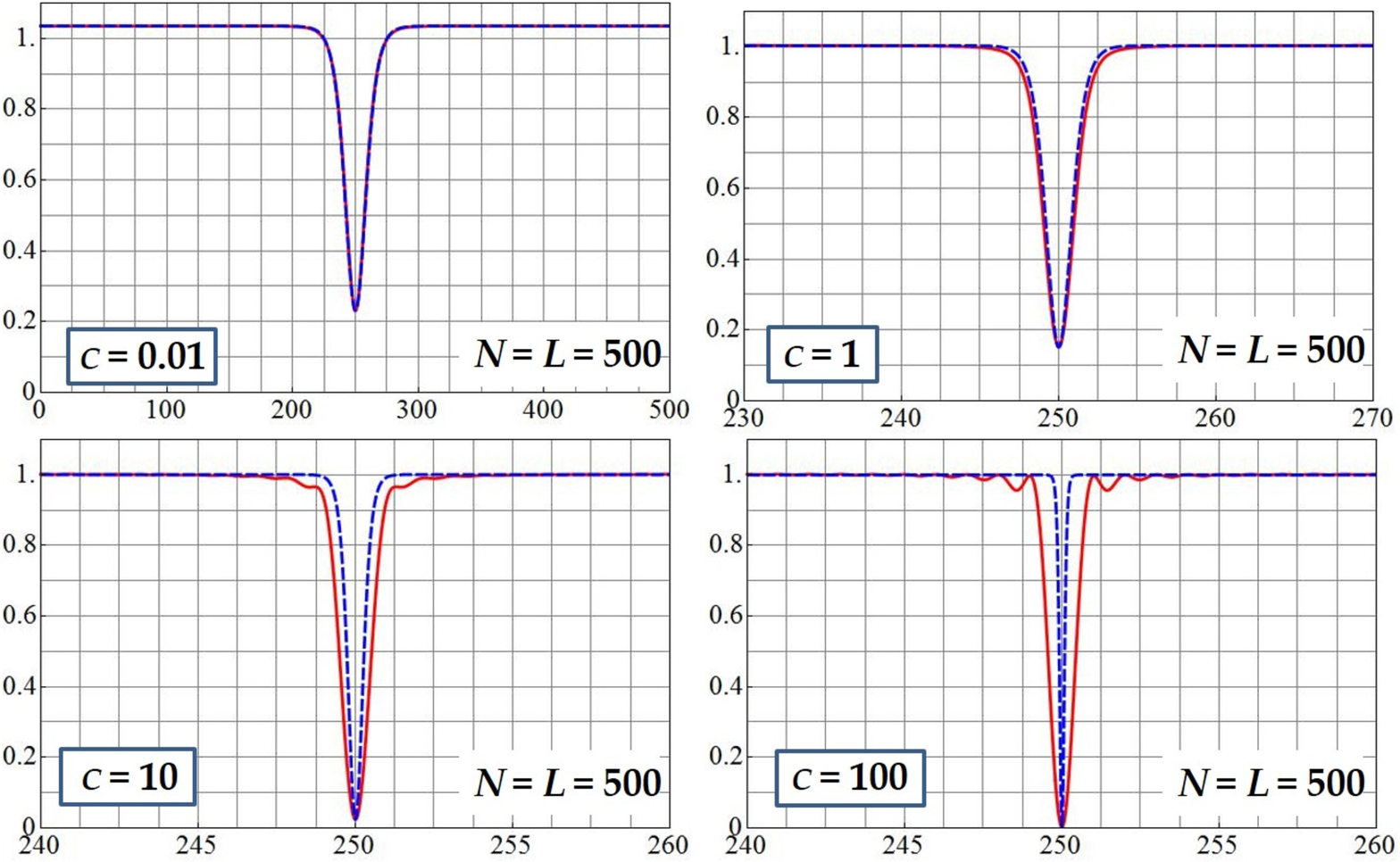}
\caption{
Density profile of quantum dark soliton 
with $N=L=500$, i.e. $\bra X, N|\hat{\rho}(x)|X, N\ket$ versus $x$, 
is shown by a red solid line to each value of $c$ with $c=0.01, 1, 10$ and 100. 
The density profile of classical dark soliton $|\mfperi(x)|^2=\rho(x-L/2)$ 
with parameters shown in Table \ref{density_table500-T2} 
is plotted with a blue broken line to each of $c$  with $c=0.01, 1, 10$ and 100. }
\label{density500}
\end{figure}

\begin{table} \begin{tabular}{c|cccc}
$c$ &  $k$ &$\mu/2nc$  &$v_c/2\sqrt{nc}$  &$v/v_c$ \\
\hline
0.01&$0.680833$       &1.19680 &1.75322 &0.985292\\
1  &$1 \! - \! 1.52994\!\times\!\!10^{-8 }$ &1.05335 &0.958113 &0.451252\\
10 &$1 \! - \! 3.64668\!\times\!\!10^{-27}$ &1.02180 &0.979219 &0.250125\\
100 &$1 \! - \! 4.97633\!\times\!\!10^{-85}$ &1.00677 &0.993330 &0.233027 \end{tabular} 
\caption{Parameters for the density profiles of the classical dark solitons with $L=20$  plotted by blue broken lines in Fig. \ref{density20} 
and in the first and the second columns of Fig.  \ref{density_dynamics}. } 
\label{density_table20-T1} 
\end{table} 

\begin{table} \begin{tabular}{c|cccc}
$c$ &  $k$ &$\mu/2nc$  &$v_c/2\sqrt{nc}$  &$v/v_c$ \\
\hline
0.01&$1 \! - \! 2.73570\!\times\!10^{-19 }$&1.01418& 0.987716&0.531055\\
1  &$1 \! - \! 3.12828\!\times\!10^{-200 }$&1.00186& 0.998163&0.395100\\
10 &$1 \! - \! 1.26650\!\times\!10^{-678 }$&1.00097& 0.999030&0.156986\\
100 &$1 \! - \! 2.22915\!\times\!10^{-2169}$&1.00037& 0.999629&0.0468324 \end{tabular} 
\caption{Parameters for the classical dark solitons with $L=500$ plotted by blue broken lines in Fig. \ref{density500}  and in the third and the fourth columns of Fig.  \ref{density_dynamics}. } 
\label{density_table500-T2} 
\end{table}

We now show that the density profile of quantum dark soliton is in good agreement 
with that of classical dark soliton, i.e.  the profile of the square amplitude of the complex scalar field for the corresponding classical dark soliton: $|\mfperi(x)|^2$ versus coordinate $x$. 
In particular, the quantum and classical density profiles are consistent with each other  
in the weak coupling case such as $c=0.01$. 

In Figs. \ref{density20} and \ref{density500} for the cases of $N=L=20$ and $N=L=500$, respectively, the density profile of quantum dark soliton is plotted with a red solid line to each value of the coupling constant $c$ with $c=0.01, 1, 10$ and $100$. 
The density profiles of classical dark soliton, $|\mfperi(x)|^2$, are plotted with blue broken lines in Figs. \ref{density20} and \ref{density500} for $N=L=20$ and $N=L=500$, respectively, to the four values of $c$ with $c=0.01, 1, 10$ and $100$. 
In the weak coupling case such as $c=0.01$ the density profiles of quantum dark soliton are consistent with those of classical dark solitons, as shown in the upper left panels of Figs. \ref{density20} and \ref{density500}. We thus suggest that the consistency of the density profile 
of quantum dark soliton with that of  classical dark soliton should be exact in the weak 
coupling limit of $c\to0$. 
 
We depict the density profile of classical dark soliton by shifting the profile of square amplitude 
$\rho(x)$ derived in section 3 by $L/2$ in the $x$ direction, as shown in  Figs. \ref{density20} and \ref{density500}.   Here we remark that in section 3 we have put the density notch of any classical  dark soliton at the origin by assuming relations \eqref{eq:conditions}. 
On the other hand, we observe that if we set the parameter $X$ as $X=0$ in \eqref{eq:XN},  
it follows from formula \eqref{eq:DP-formfactor} that  the notch is located at $x=L/2$ in the density profile of quantum dark soliton.  We therefore shift the profiles of classical dark soliton by $L/2$ as $|\mfperi(x)|^2= \rho(x-L/2)$. 

The numerical estimates for the parameters of the classical dark solitons, 
i.e. elliptic modulus $k$, chemical potential $\mu$, soliton velocity $v$ and  critical velocity $v_c$ for the four values of the coupling constant $c$ (i.e. $c=0.01, 1, 10$ and 100) 
are listed in Table \ref{density_table20-T1} and Table \ref{density_table500-T2} for the cases of $N=L=20$ and $N=L=500$, respectively. They are obtained by applying the method of 
section 3.2.4 to the density profiles of quantum dark soliton. 

\subsubsection{Time evolution of the density profile of quantum dark soliton}

Let us now study the dynamics of a quantum dark soliton state and compare it with that of a classical dark soliton. We shall show that the time evolution of the density profile of quantum dark soliton is consistent with that of the corresponding classical dark soliton in the weak coupling case. 

Making use of the determinant formula of the form factors (i.e. the matrix elements) \eqref{eq:Slavnov_density} 
we evaluate numerically the expectation values of the local density operator 
for the quantum dark soliton state at time $t$ and position $x$ by 
\begin{align}
\bra X, N|\hat{\rho}(x,t)|X, N\ket
&=
\frac{1}{N}\sum^{N-1}_{p,p'=0}
\exp\[2 \pi \i (p-p') \frac{(x-X)}{L} \]
\exp\[-\i(E-E')t\]
\bra P', N|\hat{\rho}(0,0)|P, N\ket . 
\end{align}
Here we consider the quantum field operator $\hp(x,t)$ at position and time coordinates 
$x$ and $t$, respectively, and denote the local density operator by $\hat{\rho}(x,t) = \hpd(x, t) \hp(x,t)$. 

\begin{figure}
\includegraphics[width=1.00 \columnwidth]{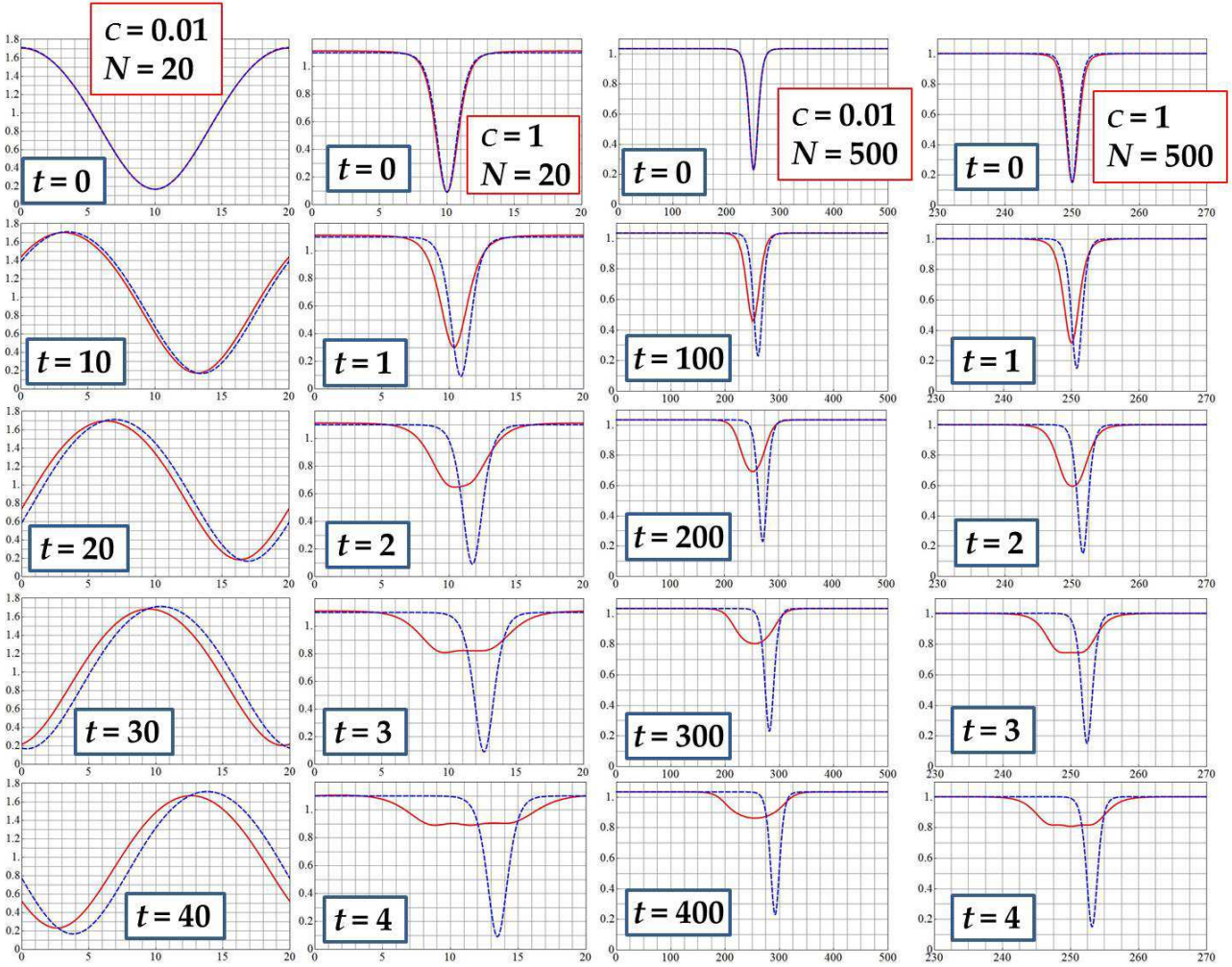}
\caption{
Snapshots at times $t$ for the density profiles of quantum dark soliton with $N=L=20$ for $c=0.01$ and $c=1$ are depicted by red solid lines in the first and the second columns from the left, respectively, and those of $N=L=500$ for $c=0.01$ and $c=1$ in the third and the fourth columns, respectively. 
Snapshots at times $t$ for the density profile of classical dark soliton 
$|\mfperi(x-vt)|^2=\rho(x-vt-L/2)$ with $L=20$ are plotted by blue broken lines for $c=0.01$ and $c=1$ in the first and the second columns from the left, respectively, and those of $L=500$ are plotted by blue broken lines for $c=0.01$ and $c=1$ in the third and the fourth columns from the left, respectively. 
The estimates of the soliton parameters for $L=20$ and $L=500$ are given in Table \ref{density_table20-T1} and Table \ref{density_table500-T2}, respectively. }
\label{density_dynamics}
\end{figure}

We now show the time evolution of the density profile of quantum dark soliton and that of  classical dark soliton, explicitly in Fig. \ref{density_dynamics}. Here we remark that the density profile at time $t$ is given by the graph of the expectation value of the local density operator 
$\hat{\rho}(x,t)$ in the quantum dark soliton state  (i.e.  $\bra X, N|\hat{\rho}(x,t)|X, N\ket$) at given time $t$ plotted against the position coordinate $x$. 

When $N=L=20$ the snapshots with $c=0.01$ and $c=1$ at times $t$ are plotted for the density profile of quantum dark soliton by red solid lines in the first and second columns of panels of Fig. \ref{density_dynamics} from the left, respectively. 
Here we remark that each column consists of five panels in Fig. \ref{density_dynamics}. 
In the first column the snapshots are  taken at $t=0, 10, 20, 30$, and $40$, 
while in the second column at $t=0, 1, 2, 3$ and $4$. 
The density profiles of classical dark soliton 
are plotted by blue broken lines for the cases of $c=0.01$ and $c=1$ in 
the first and  second columns of Fig. \ref{density_dynamics}  from the left, respectively. 
Here we recall that  the density profile of classical dark soliton at time $t$ is given by 
the square amplitude of the classical  scalar field $|\mfperi(x-vt)|^2 = \rho(x-vt-L/2)$.  
The numerical estimates of the parameters are given in Table \ref{density_table20-T1}. 

It is clear that in the weak coupling case of $c=0.01$ and $N=L=20$ 
the density profile of quantum dark soliton at time $t$ and that of  
classical dark soliton $|\mfperi(x-vt)|^2$ move together dynamically for a rather long period of time at least such as up to $t=40$ or $50$. It still keeps the initial form of the density profile in the snapshot at $t=40$, as shown in Fig. \ref{density_dynamics}. On the other hand, the density profile with $c=1$ 
collapses in a much shorter period of time such as $t=5$.  

When $N=L=500$ the snapshots with $c=0.01$ and $c=1$ at times $t$  are plotted 
for the density profile of quantum dark soliton 
by red solid lines in the third and fourth columns of panels of Fig. \ref{density_dynamics} from the left, respectively. 
In the third column the snapshots are taken at $t =0, 100, 200, 300$ and 400, 
while in the fourth column at $t = 0, 1, 2, 3$, and 4. 
The density profiles of classical dark soliton at time $t$, $|\mfperi(x-vt)|^2=\rho(x-vt-L/2)$, are plotted by blue broken lines for $c=0.01$ and $c=1$ 
in the third and fourth columns of panels of Fig. \ref{density_dynamics} from the left, 
respectively. Here, numerical estimates of the parameters are given in Table \ref{density_table500-T2}. 

We observe in the third column of Fig. \ref{density_dynamics} from the left 
that the density  profile of quantum dark soliton for $c=0.01$ moves extremely slowly 
in time evolution and keeps overlapping with  
the density profile of classical dark soliton up to $t=100$.  Then, the density notch collapses slowly over a long period of time,  and the density profile relaxes to a nonzero flat profile. 

\subsubsection{Density profile of the free-fermionic quantum-dark-soliton state
}

In the large coupling case such as $c=100$, the density profile of quantum dark soliton 
shows ripples at the shoulders of the notch, as shown in  Figs. \ref{density20} and \ref{density500}.  We now show that they are described 
in terms of the Dirichlet kernel. They correspond to the Friedel oscillations 
in the strong coupling limit. 

Let us recall the 1D free fermions in section 2.4.  
We define a state $| \Psi \rangle$  which is analogous to the quantum dark soliton state \eqref{eq:XN}  by 
\begin{equation} 
|\Psi \rangle = \frac 1 {\sqrt{N}} \sum_{\beta=1}^{N} (-1)^{\beta-1} 
\prod_{\ell=1; \ell \ne \beta}^{N+1} a_{\ell}^{\dagger} |0 \rangle \label{eq:def-psi} \, . 
\end{equation} 
The number of fermions in $|\Psi \rangle $ is given by $N$. 
Here we remark that the state $|\Psi \rangle $ is given by the sum over $N$ one-hole states, while the  state $| \Phi \rangle$ defined in \eqref{eq:stateM} consists of the sum over $N+1$ one-hole states. 

By applying the fermionic field operator $\psi(x)$ (\ref{eq:field}) to the state $|\Psi \rangle$  we have 
\begin{equation} 
\psi(x) | \Psi \rangle =  \frac 1 {\sqrt{LN}} 
\left\{ \sum_{\alpha < \beta < N+1} (-1)^{\alpha + \beta } 
\left( e^{ i k_{\alpha} x }  - e^{i k_{\beta} x }  \right) 
\prod_{\ell=1; \ell \ne \alpha, \beta}^{N+1} a^{\dagger}_{\ell} | 0 \rangle  
+ e^{i k_{N+1}x} \sum_{\beta=1}^{N} (-1)^{N+\beta} \prod_{\ell=1; \ell \ne \beta}^{N+1}  
a_{\ell}^{\dagger} | 0 \rangle \right\} \, .  \label{eq:phiPsi}
\end{equation} 
By taking the Hermitian conjugate of \eqref{eq:phiPsi} with $x$ replaced by $y$ and 
then by multiplying the original one \eqref{eq:phiPsi} by it,    
we derive the one-particle density matrix for the state $|\Psi \rangle$, rigorously.  
The result is given by 
\begin{equation}  
\langle \Psi  | \psi^{\dagger}(y) \psi(x) | \Psi \rangle  
 =  {\frac 1 L} \sum_{\alpha=1}^{N+1} e^{i k_{\alpha} (x-y)} 
- {\frac L {N}}  \left( {\frac 1 L} \sum_{\alpha=1}^{N} e^{i k_{\alpha} x  } \right) 
  \left( {\frac 1 L} \sum_{\beta=1}^{N} e^{- i k_{\beta} y  } \right)  \, .  
\label{eq:1PDMt0}
\end{equation}  

For an illustration, let us consider the odd-$N$ case with $N= 2N_c +1$ for an integer $N_c$  
and assume the set of momenta  $\{k_1, k_2, \ldots, k_N \} = \{ 2\pi n/L | \,  n = 0, \pm 1, \pm 2, \ldots, \pm N_c\}$. It is for the ground state of $N$ free fermions. Here, we put 
$k_{N+1}=2 \pi (N_c+1)/L$.   
In terms of the Dirichlet kernel $D_{N_c}(x)$ in  (\ref{eq:truncatedsum}) 
we have from \eqref{eq:1PDMt0}   
\begin{equation} 
\rho_{\Psi}(x)  =  {\frac {N+1} L} - {\frac L N}  \left( D_{N_c}(x) \right)^2   \, . 
\label{eq:oddNrhoFF}
\end{equation}
In terms of $N$,  both for the odd and even $N$ cases, we can show that the density profile of 
$| \Psi \rangle$ is given by 
\begin{equation}
\rho_{\Psi}(x) =  {\frac {N+1} L} - {\frac 1 {L N}} {\frac {\sin^2 (N \pi x/L)} {\sin^2(\pi x/L)}}  \, .  
\label{eq:DPFF}
\end{equation}
It is easy to show that the local density becomes very small but nonzero at $x=0$.
\begin{equation}     
\rho_{\Psi}(0) = 1/L . \label{eq:rho_minimum} 
\end{equation} 

Interestingly, in the large coupling case such as $c=100$, 
the density profiles of quantum dark soliton shown in Figs. \ref{density20} and \ref{density500} coincide with those of the 1D free fermions \eqref{eq:DPFF} over the whole region.  We thus have a conjecture that when we send the coupling constant $c$ to infinity, i.e. in the TG limit, the density profile of quantum dark soliton in the 1D Bose gas approaches that of the 1D free fermions. 
%

Let us now derive the time evolution of the density profile 
for the free-fermionic quantum-dark-soliton state $|\Psi \rangle$.  
The Hamiltonian of the 1D free fermions is given by  
\begin{equation} 
H = \sum_{\alpha=1}^{\infty}  \omega_{\alpha} a^{\dagger}_{\alpha} a_{\alpha} \, . 
\end{equation}
When we make connection to the 1D Bose gas, 
we assume the quadratic dispersion relation: $\omega_{\alpha} = k^2_{\alpha}$.  
We now define the initial state $| \Phi(0) \rangle$  by  $| \Phi(0) \rangle = | \Phi \rangle$
and introduce its time evolution by applying the time evolution operator $e^{-iHt}$    
\begin{equation} 
|\Phi(t) \rangle = e^{-iHt} \, | \Phi(0) \rangle  \, . 
\end{equation}
We define function $F_N(x,t)=F_N(x,t; \{ k_1, \cdots, k_N \})$ by 
\begin{equation}
 F_N(x,t) =   {\frac 1 L} \sum_{\alpha=1}^{N} e^{i (k_{\alpha} x - \omega_{\alpha} t) } \, . 
\end{equation}
Similarly as the case of the one-paticle density matrix at the initial time,   
we can show that the one-particle density matrix  
for the state $| \Phi(t) \rangle$ constructed by the 
sum over one-hole states  is given by 
\begin{equation}  
\langle \Psi (t) | \psi^{\dagger}(y) \psi(x) | \Psi(t) \rangle  =  F_{N+1}(x-y, 0) 
- {\frac L {N}}  F_{N}(x,t) \overline{F_{N} (y, t)} \,  .  \label{eq:1PDM-Psi-time}
\end{equation} 
Here $F_{N+1}(x,t)=F_{N+1}(x, t; \{k_1, \ldots,  k_{N+1} \}) $ denotes the exponential sum 
for $N+1$ momenta.   
For an illustration,  we consider the odd $N$ case with the set of 
momenta for the local density \eqref{eq:oddNrhoFF}. We have   
\begin{equation}
F_N(x,t) 
= {\frac 1 L} \sum_{n=-N_c}^{N_c}  \exp\left( 2 \pi i n x /L - i t \left(4 \pi^2 n^2 /L^2 \right) \right) . \label{eq:FNxt}
\end{equation}

In the large coupling case such as $c=100$, the time evolution of the density profile of quantum dark soliton looks quite similar  to that of the free-fermionic quantum-dark-soliton state $| \Psi \rangle$ calculated by (\ref{eq:1PDM-Psi-time}) with  (\ref{eq:FNxt}).

\subsubsection{Gaussian weighted sum leading to a smooth density profile}

We recall that in the large coupling case such as $c=100$, the density profiles of quantum dark soliton  
show ripples at the shoulders of the notches, as shown in 
Figs. \ref{density20} and \ref{density500}.  They correspond to  the oscillations  
in the Dirichlet kernel.  By modifying the construction \eqref{eq:XN} of quantum dark soliton  
with the Gaussian weights 
we can remove the ripples in the density profile of quantum dark soliton so that  it  consists of only smooth curves.  In Appendix B we shall show that a smooth Gaussian profile is obtained by introducing the Gaussian weight to the sum.  

However, in the large coupling case, the width of the corresponding classical dark soliton can be much smaller than the width of the notch in the density profile of quantum dark soliton. Let us recall relation (\ref{eq:a1}) in section 3. Here, the parameter $a_1$ gives the smallest value of the soliton density $\rho(x)$. 
In the large coupling case such as $c=100$, it is given by $1/L$, as shown in   
\eqref{eq:rho_minimum}. Here, we set $n=N/L=1$. 
When the modulus $k$ is close to 1, the complete elliptic integrals 
$K$ and $E$ are given by $K= O ( \log(1/ p))$ and $E= O (1)$ 
as shown in the asymptotic behavior \eqref{eq:asymptotic}  
in section 3.3.
It follows from relation (\ref{eq:a1}) that for large  $c$ we have 
\begin{equation} 
1/L \approx 1 - 4K^2/(c L^2 ) . 
\end{equation} 
Therefore, we have the approximate evaluation of parameter $b$ defined by $b=K/L$ 
in the case of large $c$: $b= K/L \approx \sqrt{c}/2$. 
Considering the $x$-dependence of the square amplitude (\ref{eq:rho-EK})
and that of the phase (\ref{eq:phase-theta}) of a classical dark soliton 
under the PBCs  we define the correlation length $\xi$ 
by $\xi= 1/(2b)$.  Therefore we have  
\begin{equation}
\xi \approx 1/\sqrt{c} \quad (c \gg 1) .  
\end{equation}
It can be smaller than the standard deviation of the position operator in the position-momentum uncertainty relation, as the coupling constant $c$ becomes very large. Consequently, even the modified Gaussian density profile of quantum dark soliton does not fit to the density profile of classical dark soliton.

\subsection{Profiles of the square amplitude and the phase of the quantum field operator}
\subsubsection{Formula for the off-diagonal matrix elements of the quantum field operator }

Let us consider the matrix element of the quantum field operator $\hp(x)$
between the quantum dark soliton states with $N$ particles, $| X, N \ket$, 
and that of $N-1$ particles, $|X^{'}, N-1 \ket$. We define the symbol $\qs(x)$ by  
\begin{align}
&\qs(x):=
\bra X', N-1|\hp(x)|X, N \ket
\nn\\&
=
\frac1{\sqrt{N(N-1)}}
\sum^{N-1}_{p=0}
\sum^{N-2}_{p'=0}
\exp\[2 \pi \i (p-p')\frac{x}{L}\]
\exp\[-2 \pi \i \(\frac{pX}{L}-\frac{p'X'}{L}\)\]
\bra P', N-1|\hp(0)|P, N\ket, \label{eq:matrix}
\end{align}
where $P=2\pi p/L$ and $P'=2\pi p'/L$ denote the total momenta 
of the normalized Bethe eigenstates in the type II branch 
$|P, N\ket$ and $|P', N\ket$, respectively. 
Here we remark that the two quantum states have different numbers of particles such as $N$ and $N-1$ but they share the same system size $L$. 
Hereafter we consider only the case of $X=X'=0$. 
%
The matrix element $\bra P', N-1|\hp(0)|P, N\ket$ are evaluated effectively by the determinant formula for the 
norms of Bethe eigenstates \cite{GK} and that for the form factors of the field operator 
\cite{Slavnov, Kojima, Caux-Calabrese-Slavnov2007} as
\begin{align}
&\bra P', N-1|\hp(0)|P, N\ket
\nn\\&
=(-1)^{N(N+1)/2+1}
\(
\prod^{N-1}_{j=1}
\prod^N_{\ell=1}
\frac{1}{k'_j-k_\ell}\) 
\( 
\prod^N_{j>\ell}k_{j,\ell}\sqrt{k_{j,\ell}^2+c^2} 
\)
\( 
\prod^{N-1}_{j>\ell}\frac{k'_{j,\ell}}{\sqrt{(k'_{j,\ell})^2+c^2}} 
\)
\frac{\det \widehat{U}(k,k')}{\sqrt{\det G(k)\det G(k')}}, 
\label{eq:Slavnov_field}
\end{align}
where the quasi-momenta $\{k_1,\cdots,k_N\}$ and $\{k'_1,\cdots,k'_{N-1}\}$ 
lead to the eigenstates $|P, N\ket$ and $|P', N-1\ket$, respectively. 
%
%
Here we have employed the abbreviated symbols 
$k_{j,\ell}:=k_j-k_\ell$ and $k'_{j,\ell}:=k'_j-k'_\ell$. 
Here we recall that the kernel ${\hat K}(k)$ is given by ${\hat K}(k)=2c/(k^2+c^2)$ and 
the matrix $G(k)$ denotes the Gaudin matrix, whose $(j,\ell)$th element is 
given in \eqref{eq:Gaudin}. 
%
%
The matrix elements of the $(N-1)$ by $(N-1)$ matrix $\widehat{U}(k,k')$ are given by 
\begin{align}
\widehat{U}(k,k')_{j,\ell}&=2\delta_{j\ell}\text{Im}\[
\frac{\prod^{N-1}_{a=1}(k'_a-k_j + ic)}{\prod^N_{a=1}(k_a-k_j + ic)}\]
+\frac{\prod^{N-1}_{a=1}(k'_a-k_j)}{\prod^N_{a\neq j}(k_a-k_j)}
\({\hat K}(k_{j,\ell})-{\hat K}(k_{N,\ell})\). 
\label{eq:matrixU_field}
\end{align}

\begin{figure}
\includegraphics[width=0.9\columnwidth]{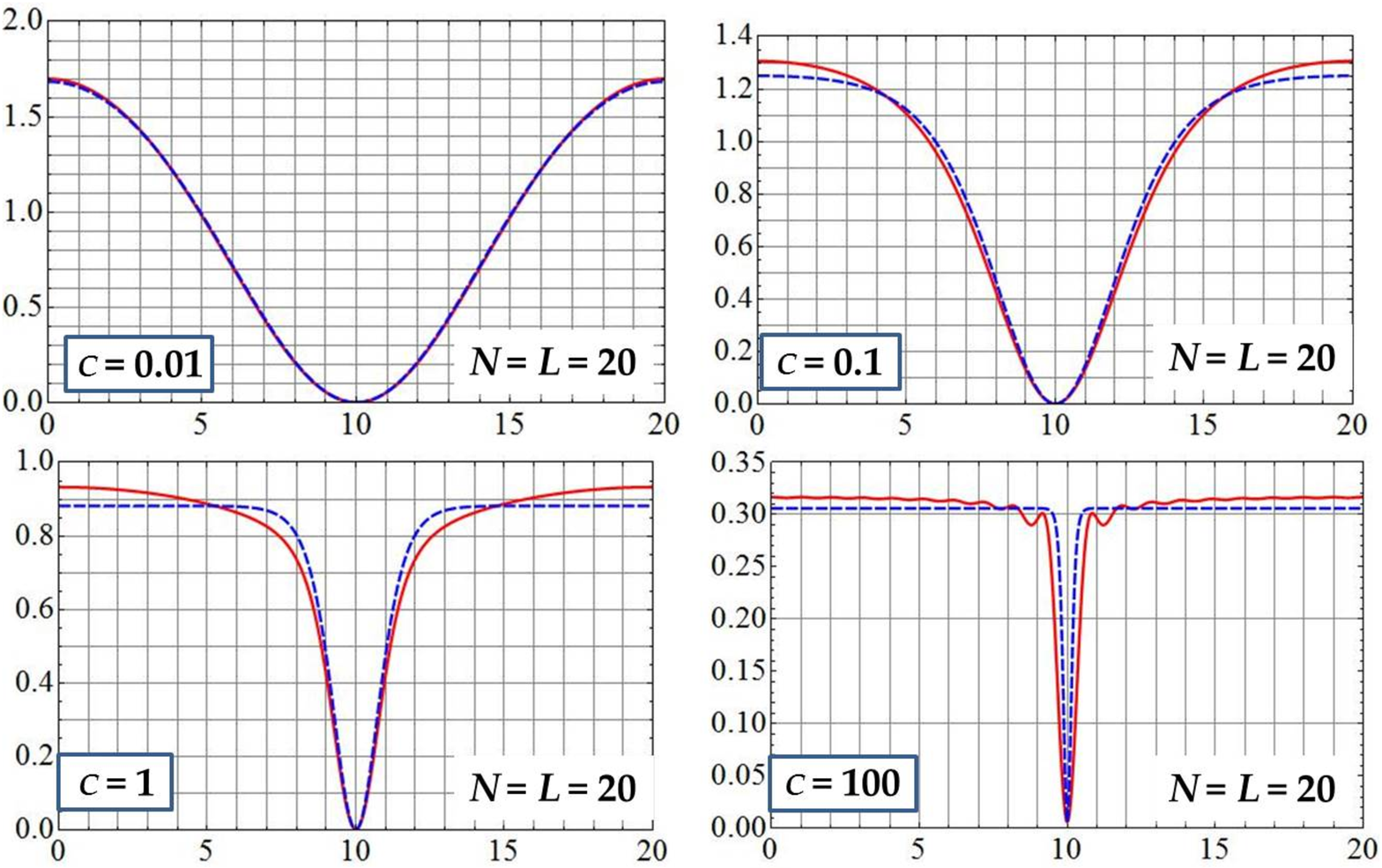}
\caption{
The profiles of the squared amplitude  $|\qs(x)|^2$ for the matrix elements of the quantum scalar field operator 
between the quantum dark soliton states
with $N=L=20$ for $c= 0.01, 1, 10$ and 100
are shown by red solid lines. 
The 
density profiles of classical dark soliton $|\mfperi(x)|^2$ 
are plotted by blue broken lines for $c= 0.01, 1, 10$ and 100. 
}
\label{field_amp20}
\end{figure}

\begin{figure}
\includegraphics[width=0.9\columnwidth]{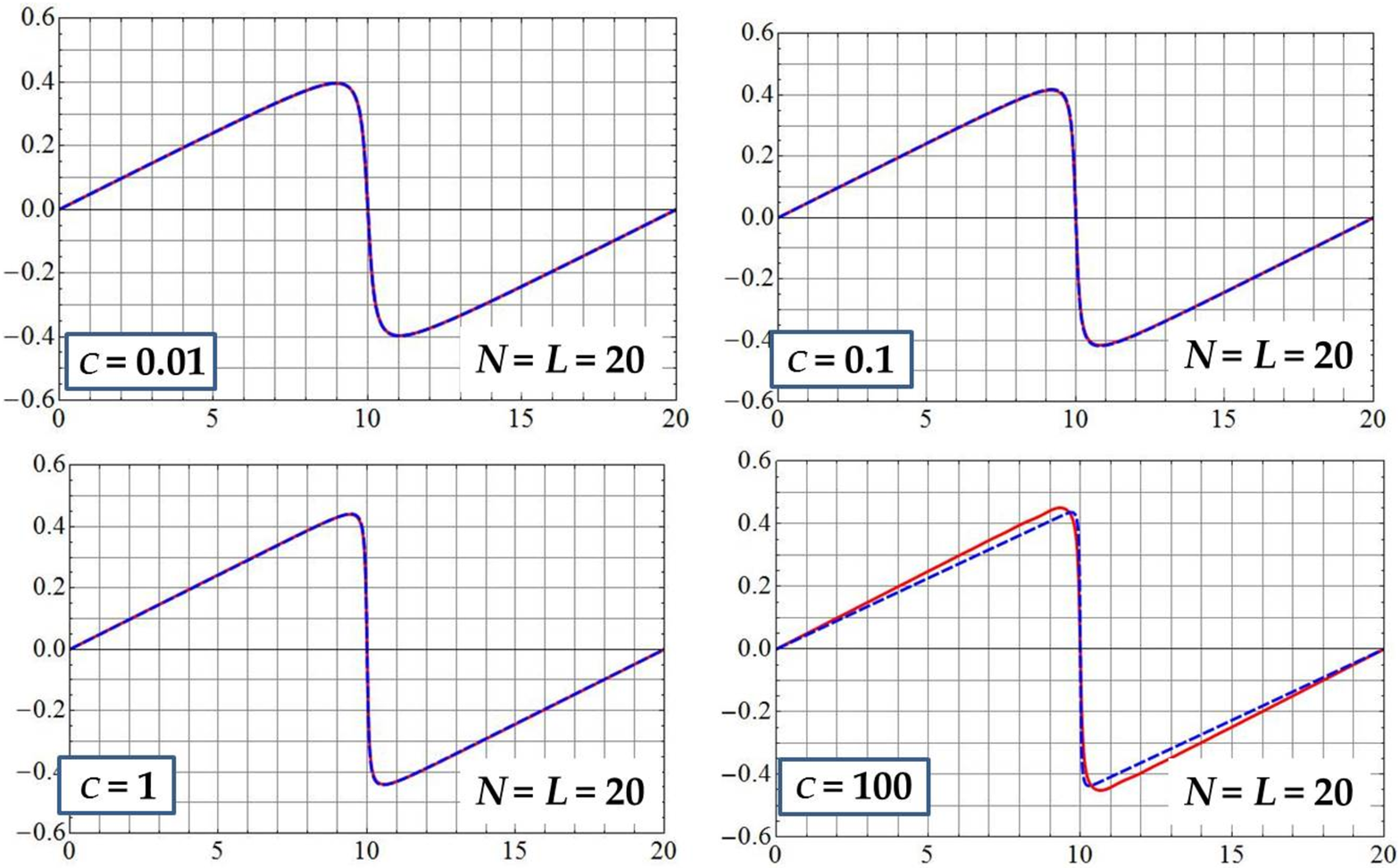}
\caption{
The phase profiles $\text{Arg}[\qs(x)]/\pi$ 
 for the matrix elements of the quantum scalar field operator between the quantum dark soliton states with $N=L=20$ are shown by red solid lines for $c= 0.01, 0.1, 1$ and 100. 
The phase profiles of classical dark soliton $\text{Arg}[\mfperi(x)]/\pi$ 
are plotted by blue broken lines for $c= 0.01, 0.1, 1$ and 100. 
}
\label{field_phase20}
\end{figure}

\begin{table} \begin{tabular}{c|cccc}
$c$ &  $k$ &$\mu/2nc$  &$v_c/2\sqrt{nc}$  &$v/v_c$ \\
\hline
0.01&0.70209                &1.41507& 1.69351 & 0.979439\\
0.1 &0.993356                &1.23980& 0.890705& 0.621731\\
1  &$1 \! - \! 5.69369\!\times\!10^{-8  }$&1.09580& 0.916354& 0.241344\\
100 &$1 \! - \! 2.29072\!\times\!10^{-47 }$&1.01408& 0.986238& 0.174410 \end{tabular} 
\caption{Parameters of the classical dark solitons for Figs.
\ref{field_amp20}, \ref{field_phase20}  and \ref{field_N20_dynamics}, where $N=L=20$. } 
\label{density_table20} 
\end{table}   

\subsubsection{Square amplitude and phase profiles for the matrix element of the quantum field operator}

Let us denote the phase argument of a given complex number 
$z=r \exp(i \theta)$ ($0 \le r$, $- \pi < \theta \le \pi$) by $\text{Arg}[z] = \theta$.  
For instance,  we shall denote the phase of the matrix element $\qs(x)$ by $\text{Arg}[\qs(x)]$.

For $N=L=20$, the profiles of square amplitude $|\qs(x)|^2$ and phase $\text{Arg}[\qs(x)]/\pi$ are plotted by red solid lines in Figs. \ref{field_amp20} and \ref{field_phase20}, respectively. We recall that they are defined for the matrix element $\qs(x)$ between the quantum dark soliton states.  The density profiles $|\mfperi(x)|^2=\rho(x-L/2)$ and phase profiles $\text{Arg}[\psi_{C}(x)]/\pi = \varphi(x-L/2)/\pi$ of classical dark soliton are shown by broken blue lines in Figs. \ref{field_amp20} and \ref{field_phase20}, respectively. The numerical estimates of the soliton parameters are given in Table \ref{density_table20}.  They are obtained by the numerical method of section 3.2.4. 
%

\begin{figure}
\includegraphics[width=0.9\columnwidth]{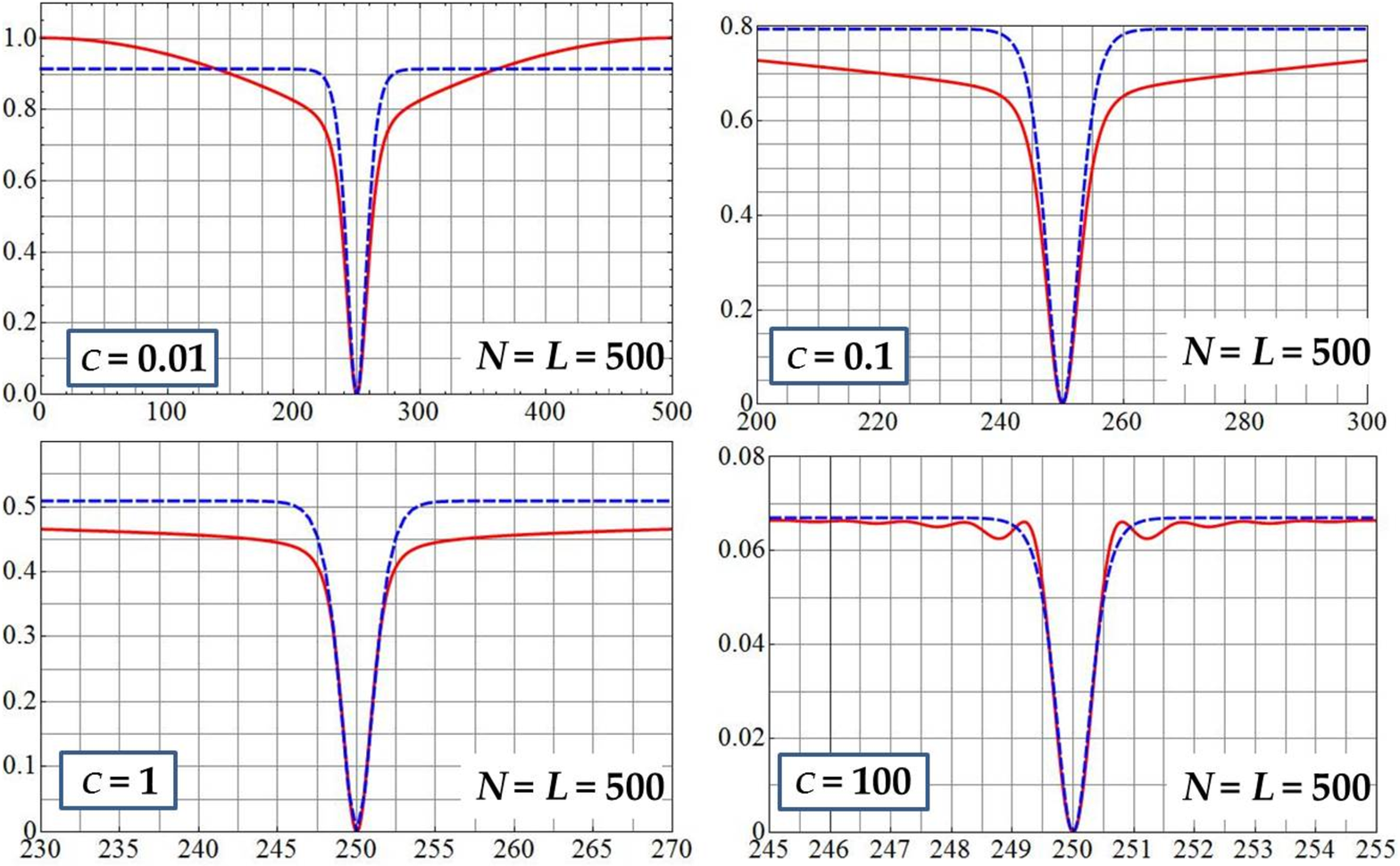}
\caption{
The profiles of the squared amplitude $|\qs(x)|^2$ 
with $N=L=500$ are shown by red solid lines 
for $c= 0.01, 0.1, 1$ and 100. The density profiles of classical  dark soliton 
$|\mfperi(x)|^2$ are plotted by blue broken lines for $c= 0.01, 0.1, 1$ and 100. 
}
\label{field_amp500}
\end{figure}
\begin{figure}
\includegraphics[width=0.9\columnwidth]{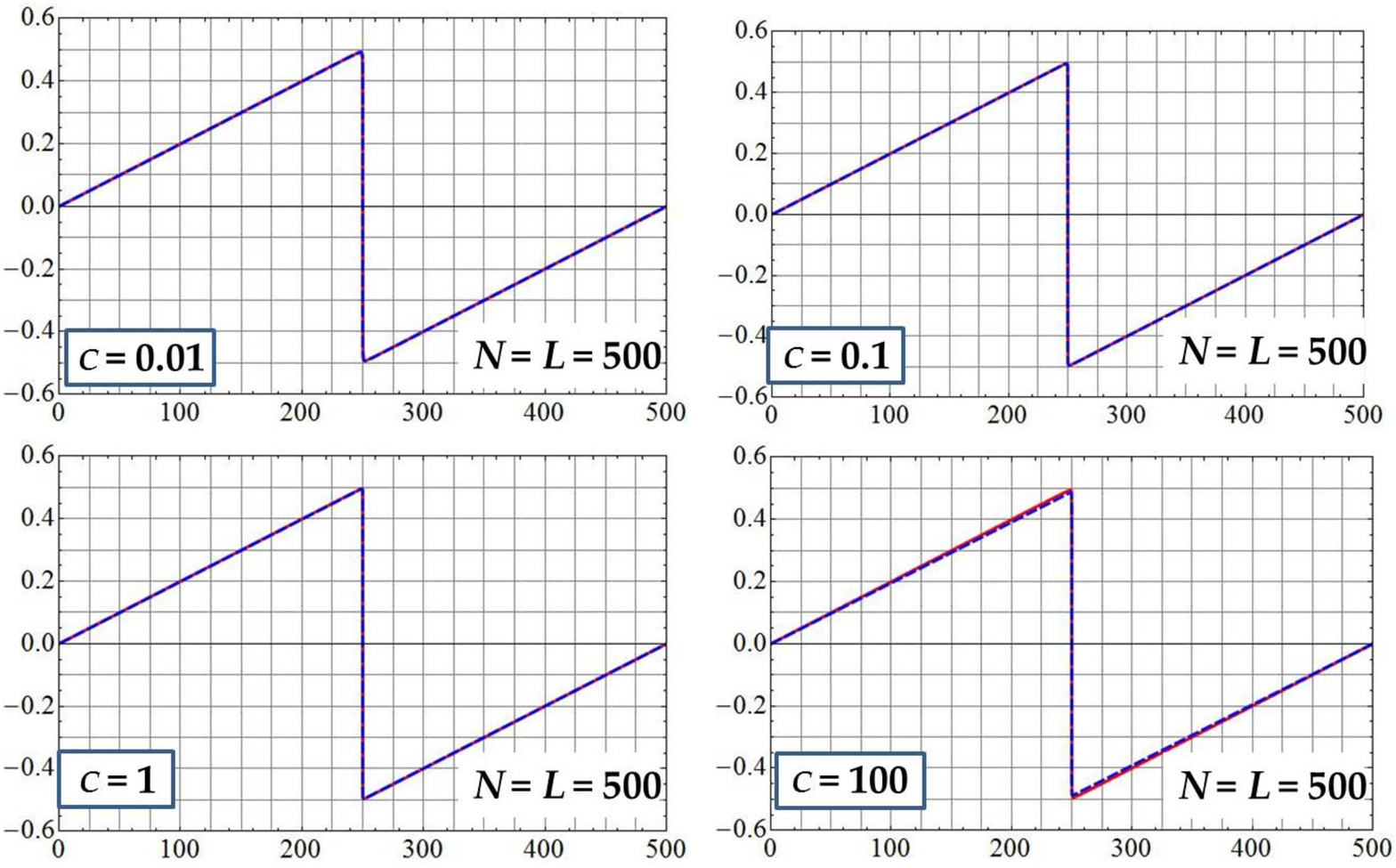}
\caption{
The profiles of the phase $\text{Arg}[\qs(x)]/\pi$ with $N=L=500$ 
are shown by red solid lines for $c= 0.01, 0.1, 1$ and 100. 
The phase profiles of classical dark soliton 
$\text{Arg}[\mfperi(x)]/\pi$ 
are plotted by blue broken lines for $c= 0.01, 0.1, 1$ and 100. 
}
\label{field_phase500}
\end{figure}

\begin{table} \begin{tabular}{c|cccc}
$c$ &  $k$ &$\mu/2nc$  &$v_c/2\sqrt{nc}$  &$v/v_c$ \\
\hline
0.01&$1 \! - \! 1.35548\!\times\!10^{-20 }$&1.04126& 0.960204& 0.0719861\\
0.1 &$1 \! - \! 4.78109\!\times\!10^{-61 }$&1.01409& 0.986069& 0.0251794\\
1  &$1 \! - \! 9.64924\!\times\!10^{-155 }$&1.00556& 0.994466& 0.0136298\\
100 &$1 \! - \! 1.03071\!\times\!10^{-561 }$&1.00146& 0.998539& 0.0363960 \end{tabular} 
\caption{Parameters of the classical dark solitons for Figs.
\ref{field_amp500},  \ref{field_phase500} and \ref{field_N500_dynamics}, where 
$N=L=500$.}  
\label{density_table500} 
\end{table}

The square amplitude profile $|\qs(x)|^2$ of quantum dark soliton 
is consistent with the density profile of classical dark soliton in the weak coupling case of $c=0.01$, as shown in the upper left panel of Fig. \ref{field_amp20}. 
Moreover, the phase profiles $\text{Arg}[\qs(x)]$ of quantum dark soliton 
perfectly agree with those of classical dark soliton for $c=0.01$ and $c=0.1$, as graphs.  
Here, no deviation is found between the quantum and classical profiles. 
We thus suggest that both the profiles of square amplitude $|\qs(x)|^2$ 
and phase $\text{Arg}[\qs(x)]$ for the matrix element $\qs(x)$ 
between the quantum dark soliton states should be consistent 
with those of classical dark soliton, respectively, in the weak coupling limit: $c\to0$. 

For $N=500$ the profiles of square amplitude $|\qs(x)|^2$ and 
phase $\text{Arg}[\qs(x)]/\pi$ are plotted in Figs \ref{field_amp500} and \ref{field_phase500}, respectively. 
The density profile  $|\mfperi(x)|^2=\rho(x-L/2)$ and phase profile 
$\text{Arg}[\psi_{C}(x)]/\pi = \varphi(x-L/2)/\pi$ of classical  dark soliton 
are shown by broken blue lines.  
The numerical estimates of the soliton parameters are given in Table 
\ref{density_table500}.  They are obtained by the numerical method of section 3.2.4 .

The phase profiles $\text{Arg}[\qs(x)]/\pi$  of the matrix element $\qs(x)$ perfectly agree with those of the corresponding classical dark solitons for the cases of $c=0.01, 0.1$ and 1, 
respectively. On the other hand, the square amplitude profile $|\qs(x)|^2$ of the matrix element $\qs(x)$ is not necessarily in very good agreement with the density profile of classical dark soliton even for the case of $c=0.01$. 
We suggest that it is due to the effect of many-body correlations in the 1D Bose gas 
such as observed in the study of the BEC fraction \cite{BEC-1DBoseGas} as mentioned in Introduction. We also suggest that if the coupling constant $c$ becomes much smaller, 
the square amplitude profiles $|\qs(x)|^2$ should be in much better agreement with the density profiles of classical dark soliton $|\mfperi(x)|^2$.

%
\subsubsection{Time evolution of the square amplitude and  phase profiles for 
the matrix element of the quantum field operator}

We now show the time evolution of the square amplitude profile and the phase profile 
of the matrix element of the bosonic quantum field operator 
with respect to the quantum dark soliton state with $N$ particles and that of $N-1$ particles.  
We denote by $\qs(x,t)$ the matrix element of the bosonic quantum field operator 
$\hp(x,t)$ between the quantum dark soliton states $|X, N \ket$ and $|X^{'}, N-1 \ket$ 
 as follows. 
\begin{equation} 
\qs(x, t) =\bra X', N-1|\hp(x, t)|X, N \ket \, . 
\end{equation} 
Here we recall that the two quantum states have different particle numbers but have the same system size $L$.

\begin{figure}
\includegraphics[width=1.0\columnwidth]{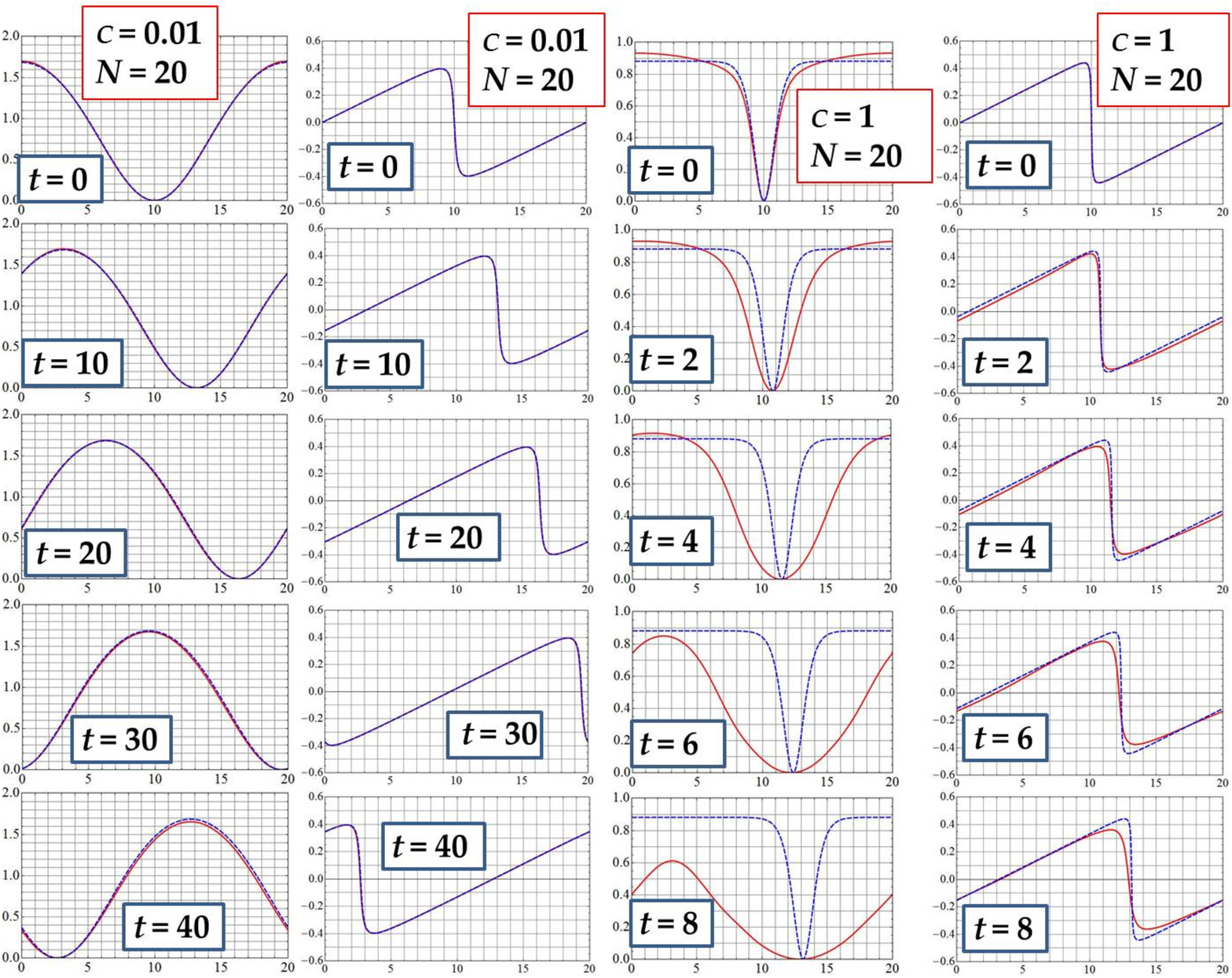}
\caption{
(i) Snapshots at times $t$ for the profiles 
of square amplitude  $|\qs(x, t)|^2$ and phase $\text{Arg}[\qs(x, t)]/\pi$ 
of quantum dark soliton with $N=L=20$ for $c=0.01$ are shown by red solid lines 
in the first and second columns from the left, respectively. 
The snapshots at times $t$ of the density profile  $|\mfperi(x-vt)|^2$ and phase profile 
$\text{Arg}[\mfperi(x-vt)]/\pi$ of classical dark soliton are plotted 
by blue broken lines in the first and second columns from the left, respectively. 
Here, the soliton parameters are given in Table \ref{density_table20}. 
(ii) Snapshots at times $t$ for the profiles 
 of square amplitude $|\qs(x, t)|^2$ and phase $\text{Arg}[\qs(x, t)]/\pi$ of  
quantum dark soliton with $N=L=20$ for $c=1$ 
are shown by red solid lines in the third and fourth columns, respectively. 
The snapshots of the density profile $|\mfperi(x-vt)|^2$ and phase profile 
$\text{Arg}[\mfperi(x-vt)]/\pi$ of classical dark soliton 
are shown by blue broken lines in the third and fourth columns 
from the left, respectively. Here, the soliton parameters are shown in 
Table \ref{density_table20}. 
}
\label{field_N20_dynamics}
\end{figure}

\begin{figure}
\includegraphics[width=1.0\columnwidth]{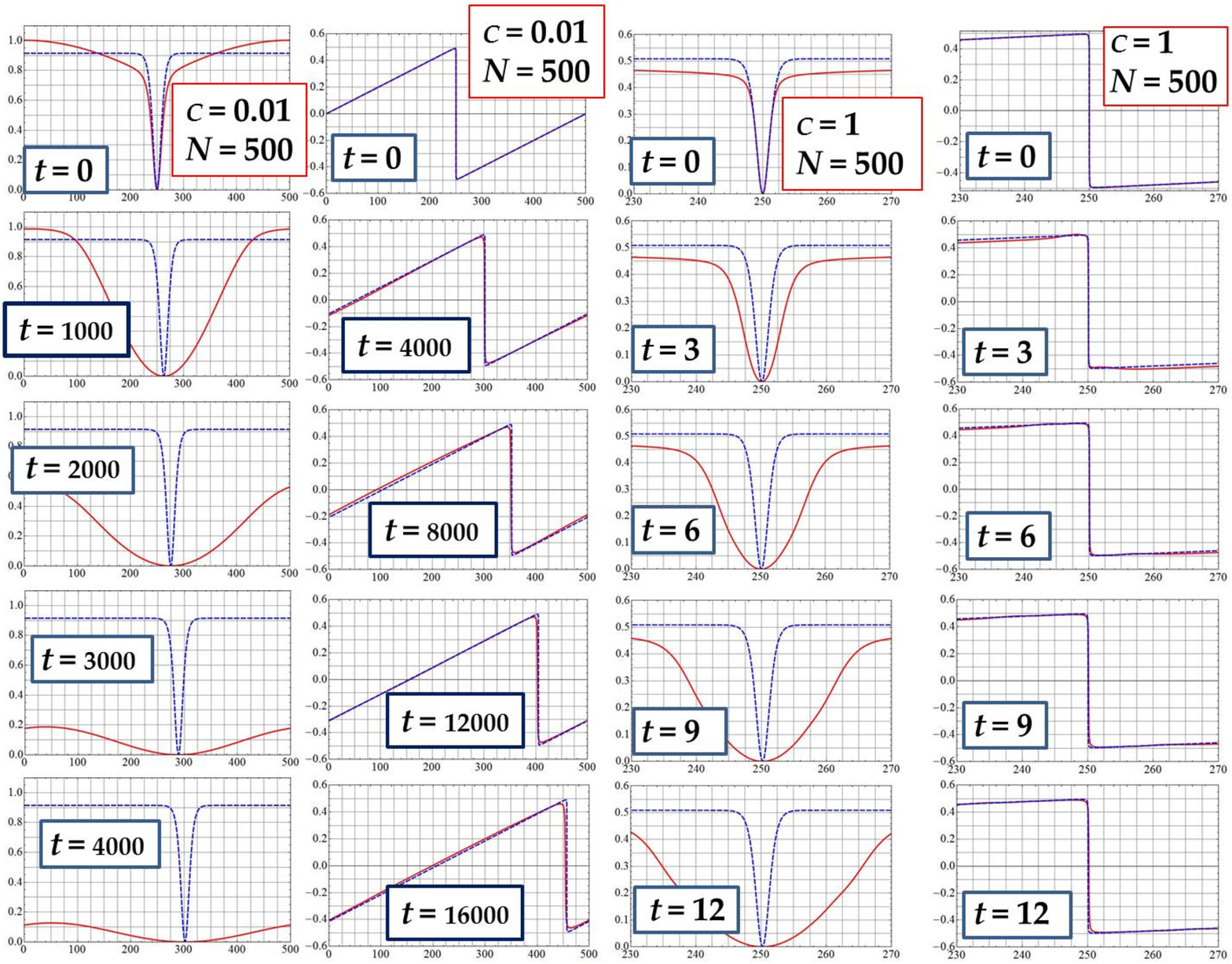}
\caption{
(i)  Snapshots at times $t$ for the profiles of square amplitude $|\qs(x, t)|^2$ and phase $\text{Arg}[\qs(x, t)]/\pi$ of quantum dark soliton with $N=L=500$ for $c=0.01$ are shown by red solid lines in the first and second columns from the left, respectively. The snapshots at times  $t$ for the density profile $|\mfperi(x-vt)|^2$ 
and  phase profile $\text{Arg}[\mfperi(x-vt)]/\pi$ 
of classical dark soliton are plotted 
by bule broken lines in the first and  second columns from the left, respectively. 
Here, the soliton parameters are given in Table \ref{density_table500}. 
(ii) Snapshots at times $t$ for  
the profiles of square amplitude $|\qs(x, t)|^2$ and phase $\text{Arg}[\qs(x, t)]/\pi$ of quantum dark soliton with $N=L=500$ for $c=1$ are shown by red solid lines in the third and fourth columns, respectively. 
The snapshots at times $t$  for the density profile $|\mfperi(x-vt)|^2$  and phase profile  
$\text{Arg}[\mfperi(x-vt)]/\pi$ of classical dark soliton are shown in the third and fourth columns from the left, respectively. Here, the soliton parameters are shown in Table \ref{density_table500}. 
}
\label{field_N500_dynamics}
\end{figure}

The snapshots at times $t$ for the profiles of the square amplitude and 
the phase of  the matrix element $\qs(x,t)$ of the bosonic quantum field operator $\hp(x,t)$ between the quantum dark soliton states $|X, N \ket$ and $|X^{'}, N-1 \ket$ 
are plotted by red solid lines for   the cases of $N=L=20$ and $N=L=500$ in  Figs. \ref{field_N20_dynamics} and  \ref{field_N500_dynamics}, respectively.  Here, we consider two values of the coupling constant such as 
$c=0.01$ and $c=1$ to each of the particle number $N$ 
in the cases of $N=20$ and 500.

The quantum and classical profiles are in good agreement in  
 many aspects of the time evolution as shown  in Figs. \ref{field_N20_dynamics} 
and \ref{field_N500_dynamics}. We now describe three aspects as follows.  
Firstly,  in the weak coupling case 
of $c=0.01$, both the time evolution of the square amplitude profile $|\qs(x,t)|^2$ and that of the phase profile $\text{Arg}[\qs(x, t)]/\pi$ are consistent with those of the corresponding classical dark soliton, $|\mfperi(x-vt)|^2$ and $\text{Arg}[\mfperi(x-vt)]/\pi$, respectively. The consistency of the quantum profiles of the square amplitude and the phase 
with the classical profiles of the corresponding classical dark soliton holds over a long period of time such as at least up to $t=40$ as shown in the first and second columns of panels 
in Fig. \ref{field_N20_dynamics} from the left. 
Secondly, the quantum profiles move with the same speed as the corresponding classical profiles, as shown in all the eight columns of panels in Figs. \ref{field_N20_dynamics} and \ref{field_N500_dynamics}.  
In the first and third columns of Figs. \ref{field_N20_dynamics} 
the positions of the notches in the square amplitude profiles $|\qs(x, t)|^2$ 
are in very good agreement with those of the corresponding classical dark solitons. 
In the second and fourth columns of Fig. \ref{field_N20_dynamics}, 
the step positions in the phase profiles $\text{Arg}[\qs(x, t)]/\pi$ 
are consistent with those of the phase profiles $\text{Arg}[\mfperi(x-vt)]/\pi$ of the corresponding classical dark solitons.  
Thirdly, as far as the phase profiles are concerned, the quantum and classical profiles 
coincide with each other very well for almost all cases 
as shown in Figs. \ref{field_N20_dynamics} and \ref{field_N500_dynamics}. In fact, 
not only the step position in the profile of $\text{Arg}[\qs(x, t)]$ 
moves together with that of the phase $\text{Arg}[\mfperi(x-vt)]$  
of the corresponding classical dark soliton, 
but also the whole profile itself coincides with the corresponding classical one.

\par \noindent 
(i) Case of $N=20$ for $c=0.01$  

With $c=0.01$ and $N=L=20$, i.e. the weak coupling case, 
for the profiles of square amplitude $|\qs(x, t)|^2$ and phase $\text{Arg}[\qs(x, t)]/\pi$ 
 of the matrix element $\qs(x,t)$, the snapshots at times $t$  are plotted by red solid lines in the first and second columns of panels of Fig. \ref{field_N20_dynamics} from the left, respectively. Here, five snapshots are shown in each column of panels for $t=0, 10, 20, 30$ and 40. 
The snapshots at times $t$ for the density profile  $|\mfperi(x-vt)|^2=\rho(x-vt-L/2)$
and phase profile $\text{Arg}[\mfperi(x-vt)]/\pi = \varphi(x-vt-L/2)/\pi$ 
of classical dark soliton are shown by blue broken lines in the first and  second columns, respectively.  
The numerical estimates of the soliton parameters are given in Table 
\ref{density_table20}.

\par \noindent 
(ii) Case of $N=20$ for $c=1$  

With $c=1$ and $N=L=20$, for the profiles
of square amplitude $|\qs(x, t)|^2$ and phase $\text{Arg}[\qs(x, t)]/\pi$  
of the matrix element $\qs(x,t)$, the snapshots at times $t$  
are plotted by red solid lines in the third and fourth columns of panels of  
Fig. \ref{field_N20_dynamics} from the left, respectively. Here, five snapshots are 
shown in each column of panels for $t=0, 2, 4, 6$ and 8. 
The density profile $|\mfperi(x-vt)|^2=\rho(x-vt-L/2)$
and phase profile $\text{Arg}[\mfperi(x-vt)]/\pi = \varphi(x-vt-L/2)/\pi$ 
of classical dark soliton  
are shown by blue broken lines in the third and fourth columns, respectively. 
The numerical estimates of the soliton parameters are given in Table 
\ref{density_table20}. 

We observe that by comparing the snapshots of $c=0.01$ with those of 
 $c=1$,  the density notch in the square amplitude profile $|\qs(x, t)|^2$ at $t=0$ decays in a shorter period of time as the coupling constant becomes larger, as shown in the first  
and third columns of Fig. \ref{field_N20_dynamics}  for $c=0.01$ and $c=1$, respectively.   
However, the position of the density notch 
moves together with that of classical dark soliton, although the notch itself becomes wider rather rapidly in time. In the phase profiles, the abrut step becomes softer and milder  
gradually in time, but the change in the phase profile is much slower than in the square amplitude profile.

\par \noindent 
(iii) Case of $N=500$ for $c=0.01$  

With $c=0.01$ and $N=L=500$,  for  the profiles of 
square amplitude $|\qs(x, t)|^2$ and phase $\text{Arg}[\qs(x, t)]/\pi$  
of the matrix element $\qs(x,t)$, the snapshots  at times $t$  
are plotted by red solid lines in the first and second columns of panels of  
Fig. \ref{field_N500_dynamics} from the left, respectively. Here, five snapshots are taken 
with much longer intervals of time than in the case of $N=L=20$ in 
Fig.  \ref{field_N20_dynamics} such as  for $t=0, 1000, 2000, 3000$ and 4000 in the 
case of the square amplitude $|\qs(x, t)|^2$ and 
 for $t=0, 4000, 8000, 12000$ and 16000 in the 
case of the phase  $\text{Arg}[\qs(x, t)]$. 
The profiles of square amplitude $|\qs(x-vt)|^2=\rho(x-vt-L/2)$
and phase $\text{Arg}[\mfperi(x-vt)]/\pi = \varphi(x-vt-L/2)/\pi$ 
are shown by blue broken lines 
in the first and second columns of Fig. \ref{field_N500_dynamics} from the left, respectively. 
The numerical estimates of the soliton parameters are given in Table 
\ref{density_table500}. 

We observe that the square amplitude $|\qs(x, t)|^2$ for the matrix element of the quantum field operator $\hp(x,t)$ approaches zero in time evolution. The decaying behavior is clear for $c=1$ as shown in the third column from the left of Figs.  \ref{field_N20_dynamics} and \ref{field_N500_dynamics}.  
In the first column of  Fig.  \ref{field_N500_dynamics} from the left, we also observe it   for  $c=0.01$ and $N=L=500$ over a very long period of time up to $t=4000$.  However,  for  the density profiles of quantum dark soliton, the density approaches a nonzero constant value in time evolution, as shown in Fig. \ref{density_dynamics}.  We suggest that the quantum dark soliton states with different particle numbers such as $N$ and $N-1$ do not have the same set of energy eigenvalues in common, so that  the off-diagonal matrix element $\qs(x, t)$ vanishes in time, finally.

\par \noindent 
(iv) Case of $N=500$ for $c=1$  

With $c=1$ and $N=L=500$,  for the profiles of square amplitude  
$|\qs(x, t)|^2$ and phase  $\text{Arg}[\qs(x, t)]/\pi$ in the matrix element 
$\qs(x,t)$, the snapshots at times $t$ are plotted by red solid lines in the third and fourth columns of panels of Fig. \ref{field_N500_dynamics} from the left, respectively. Here, five snapshots are taken 
for almost the same intervals of time as in the case of Fig. \ref{field_N20_dynamics} such as for $t=0, 3, 6, 9$ and 12 in both cases of the profiles of square amplitude  $|\qs(x, t)|^2$ and phase  $\text{Arg}[\qs(x, t)]$. The density profile $|\mfperi(x-vt)|^2=\rho(x-vt-L/2)$ and phase profile $\text{Arg}[\mfperi(x-vt)]/\pi = \varphi(x-vt-L/2)/\pi$ of classical dark soliton are shown by blue broken lines in the third and the fourth columns of Fig. \ref{field_N500_dynamics} from the left, respectively. The numerical estimates of the soliton parameters are given in Table \ref{density_table500}.

Here we remark that when we evaluate the matrix element $\qs(x,t)$ straightforwardly by making use of the form factor formula,  the phase of the matrix element $\qs(x,t)$ increases  with a constant velocity in time, and the phase profile moves downwards with a constant velocity in the animation. We suggest that it is related to the difference between the zero point energy of the eigenstates with $N-1$ particles and that of $N$ particles. In Figs. \ref{field_N20_dynamics} and \ref{field_N500_dynamics} 
we plotted the phase profiles which are subtracted by the constant shift in time.

\subsubsection{Matrix element of the fermionic field operator between the free-fermionic quantum dark soliton states}

As the quantum dark soliton states for the 1D free fermions with $N$ and $N-1$ particles 
we consider the following  
\begin{eqnarray} 
|\Psi, N \rangle & = & {\frac 1 {\sqrt{N}} }  \sum_{\beta=2}^{N+1}  (-1)^{\beta-1} 
\prod_{\ell=1; \ell \ne \beta}^{N+1} a_{\ell}^{\dagger} |0 \rangle \, ,  \nonumber \\ 
|\Psi, N-1 \rangle & = & {\frac 1 {\sqrt{N-1}} }  \sum_{\beta=2}^{N}  (-1)^{\beta-1} 
\prod_{\ell=1; \ell \ne \beta}^{N} a_{\ell}^{\dagger} |0 \rangle  \, . 
\end{eqnarray}
We calculate the matrix element of the field operator between the two states as 
\begin{equation}
\langle \Psi, N-1|  \psi(x,t) | \Psi, N \rangle  = 
{\frac {(-1)^{N} } {\sqrt{LN(N-1)}}}  \left \{ - (N-1) e^{i (k_{N+1} x - \omega_{N+1} t) } 
+ \sum_{\alpha=2}^{N} e^{i (k_{\alpha} x - \omega_{\alpha} t) } \right\}  \, . 
\end{equation}
Here, the exponential sum is given by the Dirichlet kernel. For instance, 
when we take $k_{N+1}=2 \pi /L$ and momenta $2 \pi n/L$ 
for $n=0$, -1, and $\pm 2, \pm 3, \ldots, \pm N_c$ where $N=2 N_c +1$   
we have 
\begin{equation} 
\langle \Psi, N-1|  \psi(x,t) | \Psi, N \rangle = {\frac {(-1)^{N} } {\sqrt{LN(N-1)}}}
\left\{-N \exp({\frac {2 \pi i x} L} ) + {\frac L N} D_{N_c}(x) . \right\} \, . \label{eq:matrixFF}
\end{equation}    

The square-amplitude and phase profiles for the matrix element of the fermionic field operator 
\eqref{eq:matrixFF}
have several features in common with those of the 1D Bose gas in the large coupling case 
shown in Figs.  \ref{field_amp20}, \ref{field_phase20}, \ref{field_amp500}, and \ref{field_phase500}, although they do not overlap with those of the 1D Bose gas even in the large coupling case. 
For instance, the square-amplitude profile vanishes at the origin, and it has a density notch around at the origin and small ripples at the shoulders of the notch;   the phase profiles change signs at the origin, etc..

%
%
\section{Stability of density profiles of quantum dark soliton}

\subsection{Quantum speed limit time in the weak coupling limit}

For a given quantum state we denote the expectation value of the Hamiltonian measured from the ground state energy by ${\cal E}$, and the square root of the variance of the expectation value of the Hamiltonian by $\Delta {\cal E}$. 
We define the quantum speed limit time ${\cal T}_0$ \cite{MT,ML,GLM2} by 
\begin{equation} 
{\cal T}_0 = {\rm max} \left( {\frac {\pi \hbar} {2 {\cal E}} } , \, 
{\frac {\pi \hbar} {2 \Delta {\cal E} } } \right) \, . 
\end{equation}
We remark that the quantum speed limit time gives a typical lifetime of a generic quantum state. 

Let us evaluate the quantum speed limit time ${\cal T}_0$ for the quantum dark soliton state $| X, N \rangle$ in the weak coupling limit where we send the coupling constant $c$ to zero. We solve the Bethe ansatz equations (\ref{BAE}) in the weak coupling limit. 
Here we recall that the Bethe quantum numbers for one-hole excitations are defined 
by eq. (\ref{eq:Ij-one-hole}).  
Let us consider an expansion of function $\arctan x$ which is valid 
if the absolute value of $x$ is larger than 1.  
\begin{equation} 
\arctan x = \sign(x) \left( {\frac {\pi} 2} - \sum_{n=1}^{\infty} 
\frac {(-1)^{n-1}} {2n-1} | x|^{-2n +1} \right) \quad \mbox{for} \, \, |x| > 1 . 
\end{equation} 
Here $\sign(x)= 1$ for $x> 0$ while $\sign(x)= -1$ for $x< 0$. 
Applying this expansion we show in the weak coupling case ($0 < c \ll 1 $) 
\begin{equation} 
\arctan \left( \frac {k_j- k_{\ell}} {c} \right) \simeq \pi \sign (k_j- k_{\ell}) \, . 
\end{equation} 
We have the solution of one-hole excited state $| P, N \rangle$ approximately as follows. 
\begin{align} 
k_j & =  0 \qquad \mbox{for} \quad 1 \le j \le N-p \nonumber \\ 
  & =  \frac {2 \pi} L \quad \mbox{for} \quad N-p < j \le N \, . 
\end{align} 
Therefore, the one-hole excitation $| P, N \rangle$ has the energy eigenvalue 
$E_p = p \left(2 \pi/L \right)^2$.  

Let us denote by $\langle E \rangle$ 
the expectation value of the energy eigenvalues for the quantum dark soliton state 
$| X, N \rangle$. We have $\langle E \rangle = \sum_{p=0}^{N-1} E_p/N $ and hence 
\begin{align}
\langle E \rangle
& =  {\frac {N-1} 2} \left( {\frac {2 \pi} L} \right)^2 \, . 
\end{align}
Since the ground-state energy is given by 0 in the weak coupling limit, we denote it also by ${\cal E}$.
We calculate the variance of the energy eigenvalues $| X, N \rangle$ as $\langle (E- \langle E \rangle)^2 \rangle = (N^2-1) \left({2 \pi} /L \right)^4 /12$ . 
We thus have the square root of the variance, $\Delta {\cal E}$, as 
\begin{equation} 
\Delta {\cal E} = {\frac {\sqrt{N^2-1}} {2 \sqrt{3}}} \left( \frac {2 \pi} L \right)^2 \, .
\end{equation} 
It follows that the quantum speed limit time for the quantum dark soliton state in the weak coupling limit is given by that for the inverse of the square root of variance ${\pi \hbar} /{2 \Delta {\cal E}}$ 
\begin{equation} 
{\cal T}_0 = {\frac {\sqrt{3} \, \pi \hbar } {\sqrt{N^2-1} } } \left( \frac L {2 \pi} \right)^2 \, . 
\end{equation} 
Here we remark that the quantum speed limit time ${\cal T}_0$ is constant in the weak coupling limit where only the coupling constant $c$ approaches zero while the density $n$ and the particle number $N$ are kept constant. 

\subsection{Observed collapse time of a notch in the density profile}
\begin{figure}
\includegraphics[width=0.5 \columnwidth]{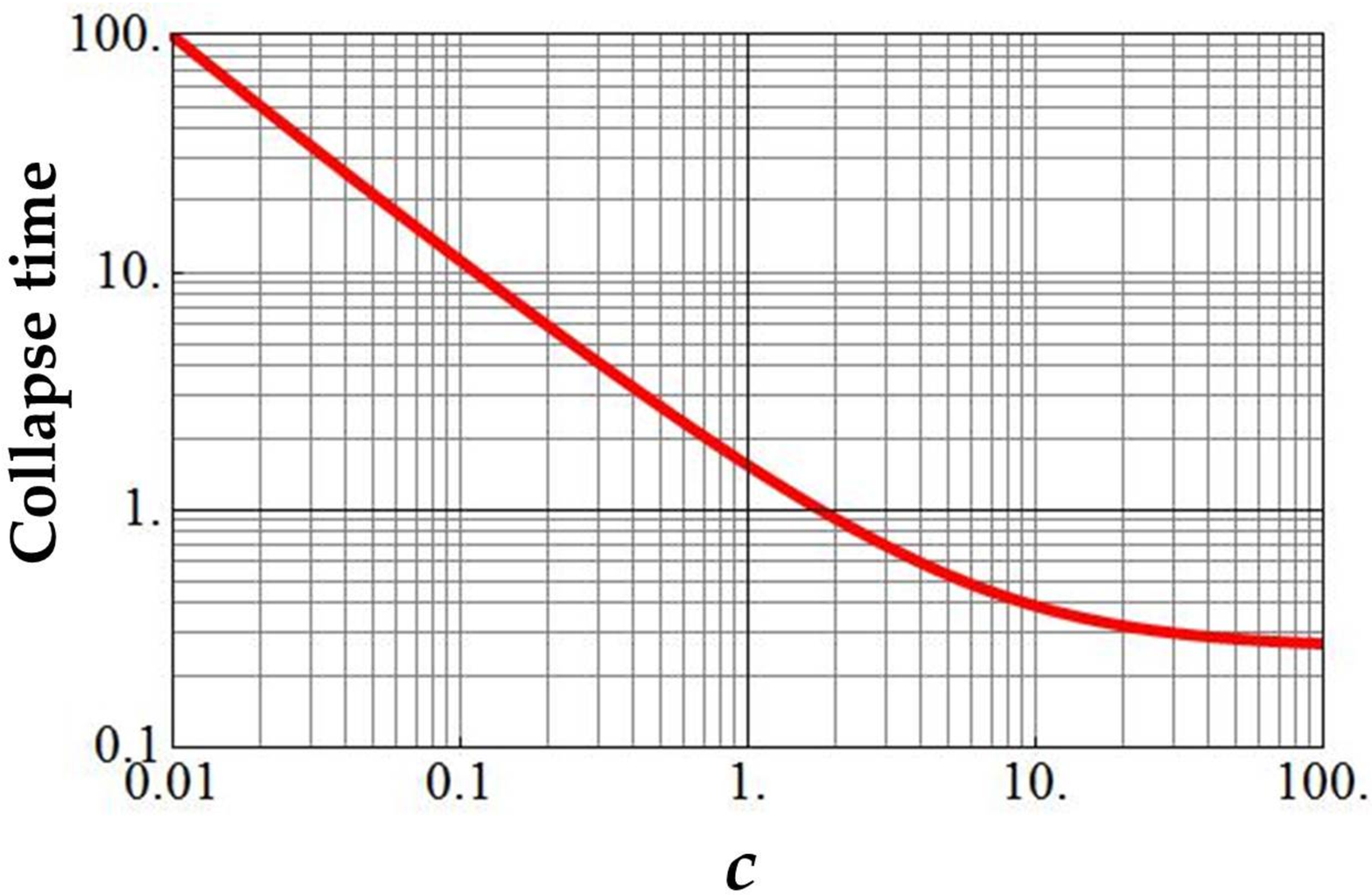}
\caption{(Color online) 
Collapse time $T_{\rm col}$ for the density profile of a quantum dark soliton state 
with $N=L=20$ versus coupling constant $c$. }
\label{RT_N20}
\end{figure}

The estimates of the relaxation time (or collapse time) for the density notch in the 
density profile of quantum dark soliton  
 for the case of $L=N=20$ are plotted against coupling constant $c$ 
in the double logarithmic scale in Fig. \ref{RT_N20}. 
By observing the time evolution of the density profile of quantum dark soliton  
for a given value of the coupling constant $c$ 
we estimated the relaxation time by the time 
when the smallest value of the density notch in the  density profile reaches the value of $0.5$. 
Here we recall that at the initial time $t=0$ the density profile of quantum dark soliton  is consistent with that of classical  dark soliton, and then the bottom level of the density profile of quantum dark soliton  increases gradually in time evolution. 

In Fig. \ref{RT_N20} we observe that the collapse time $T_{\rm col}$ increases as the coupling constant $c$ approaches zero. 
It is almost inversely proportional to the coupling constant $c$ in the weak coupling regime: 
\begin{equation} 
T_{\rm col} \simeq 1/c .
\end{equation} 
Hence we suggest that it becomes infinitely large in the weak coupling limit. 

\subsection{Stability of a notch in the density profile of quantum dark soliton} 

The quantum speed limit time ${\cal T}_0$ for quantum dark soliton states 
approaches a non-zero constant value when we send the coupling constant $c$ to zero, if we 
keep the system size $L$ and the number of particles $N$ fixed. 
In fact, in the weak coupling limit, all the solutions of the Bethe ansatz equations are given by some integral multiples of $2 \pi/L$. Hence, the quantum speed limit time of any quantum state is given by some finite value.  

On the other hand, the collapse time of the density notch in the density profiles of quantum dark soliton is approximately given by the inverse of the coupling constant, $T_{\rm col} \simeq 1/c$, in the very weak coupling case. 
Therefore, the collapse time of the density notch in the density profile of quantum dark soliton  
is much longer than the quantum speed limit time when the coupling constant $c$ is very small.

For the quantum dark soliton state with $N$ particles in the system size $L$ we have approximately 
\begin{equation} 
{\cal T}_0 \simeq \frac {\sqrt{3} \hbar} {4 \pi } \frac {L^2} {N} \, . 
\end{equation} 
In the case of $N=L=20$ we have a rough estimate: ${\cal T}_0 \simeq 2.75$. 
Here we recall that we have set $\hbar =1$.  
In Fig. \ref{RT_N20} the collapse time is given by approximately 10.0 for $c=0.1$. Thus, for 
$c < 0.1$ the collapse time of the density profile in the quantum dark soliton state 
is much longer than the quantum speed limit time. 

We conclude that the density notch in the density profile of quantum dark soliton, i.e. that of  state $| X, N \rangle$, has a much longer lifetime than a generic quantum state when the coupling constant $c$ is very small but nonzero in the 1D Bose gas interacting through the repulsive delta-function potentials. 
%
%
\section*{Acknowledgments}
The authors would like to thank Dr. A. del Campo 
for useful comments, in particular, on the quantum speed limit.  
The present research is partially supported by JSPS KAKENHI 
(Grants-in-Aid for Scientific Research) Grant No. 15K05204. 
%
%
\appendix
\setcounter{equation}{0} 
\renewcommand{\theequation}{A.\arabic{equation}}
\section{Symbols of elliptic functions}
\subsection{Elliptic integrals}
We define the elliptic integrals of the first, second and third kind, denoted by 
$F(\phi, k)$, $E(\phi,k)$ and $\Pi(n, \phi, k)$, respectively, as follows. 
\begin{align}
F(\phi,k)&
= \int^{\sin\phi}_0\frac{\d z}{\sqrt{(1-z^2)(1-k^2z^2)}} , 
\label{ei1} \\
E(\phi,k)&
= \int^{\sin\phi}_0\sqrt{\frac{1-k^2z^2}{1-z^2}}\d z , 
\label{ei2} \\
\Pi(n,\phi,k)&
= \int^{\sin\phi}_0\frac{\d z}{(1-nz^2)\sqrt{(1-z^2)(1-k^2z^2)}}
\label{ei3} \, . 
\end{align}
The complete elliptic integrals of the first, second and third kind, denoted by 
$K(k)$, $E(k)$ and $\Pi(n, k)$, respectively, are derived by substituting $\phi=\pi/2$
in (\ref{ei1}), (\ref{ei2}) and (\ref{ei3}), i.e. 
$K(k)=F(\pi/2, k)$, $E(k)=E(\pi/2, k)$ and $\Pi(n, k)= \Pi(n, \pi/2, k)$, 
respectively.  
We define the hypergeometric function
$F(a, b; c; z)$ in terms of the Gamma function $\Gamma(z)$ by 
\begin{equation}
F(a, b; c; z) = \frac {\Gamma( c)} {\Gamma(a) \Gamma(b)} 
\sum_{n=0}^{\infty} 
\frac {\Gamma(a+n) \Gamma(b+n)} {\Gamma( c+n)} \frac {z^n} {n!} \, . 
\end{equation} 
The complete elliptic integrals of the first and second kind, $K(k)$ and $E(k)$, are expressed as \cite{Copson} 
\begin{align} 
K(k)  =  {\frac {\pi} 2} F({\frac 1 2}, {\frac 1 2}; {\frac 1 2}; k^2) \, , \quad  
E(k) = {\frac {\pi} 2} F(- {\frac 1 2}, {\frac 1 2}; {\frac 1 2}; k^2) \, . \label{eq:EandK}
\end{align}
\subsection{Jacobi's elliptic functions}
Jacobi's $\sn$ function is defined by the inverse of the elliptic integral of the first kind 
\begin{align}
&u=\sn^{-1}(x,k)=\int^x_0\frac{\d z}{\sqrt{(1-z^2)(1-k^2z^2)}} . 
\label{sndef}
\end{align}
We therefore have $\sn(u,k)=x$. 
Jacobi's $\cn$ and $\dn$ functions are related to $\sn(u,k)$ as
\begin{align}
&\sn^2(u,k)+\cn^2(u,k)=1, 
\label{cndef}\\
&k^2\sn^2(u,k)+\dn^2(u,k)=1. 
\label{dndef}
\end{align}
Hereafter we shall often suppress the modulus $k$.

Jacobi's amplitude function $\am(u)$ is defined by the inverse of $x=\sin\phi$ as 
\begin{align}
&u=\sn^{-1}(x)=\sn^{-1}(\sin\phi), \nn\\
&\phi=\am(u). 
\end{align}
We express $\sn(u)$ and $\cn(u)$ as 
$\sn(u)=\sin(\am u)$ and $\cn(u)=\cos(\am u)$. 
\subsection{Integral formulas}
\subsubsection{Formulas for the complete elliptic integrals of the first and second kind}
Making the transformation of variables $z=\sn(u)$ in 
the integral expression of $E(k)$ such as \eqref{ei2} we have 
\begin{align}
\int_0^K \dn^2(u)\d u=E. 
\label{f1}
\end{align}
Due to the periodicity and the parity invariance of $\dn$ function, we have 
\begin{align}
\int_0^{2K} \dn^2(u)\d u=2E. 
\label{f2}
\end{align}
We introduce the complementary modulus $k^{'}$ by $k^{'} = \sqrt{1 - k^2}$. 
We define $K^{'}(k)$ and $E^{'}(k)$ by $K^{'}(k) = K(k^{'})$ and $E^{'}(k)=E(k^{'})$, respectively.  
Here we remark that Jacobi's elliptic functions have quarter periods $K$ and $i K^{'}$.  
We can show Legendre's relation \cite{Copson}
\begin{equation} 
E K^{'} + E^{'} K - K K^{'} = \frac {\pi} 2 \, . \label{eq:Legendre}
\end{equation}
Let us define $\tau$ and $q$ by $\tau= i K^{'}/K$ and $q=\exp(\pi i \tau)$, respectively 
\cite{Copson}. 
We have 
\begin{align}
K & =  \frac {\pi} 2 \prod_{n=1}^{\infty}\left\{(1-q^{2n})^2 (1+q^{2n-1})^4 \right\} \, , \nn \\ 
i K^{'} & =  {\frac {\pi} 2} \tau \prod_{n=1}^{\infty}\left\{(1-q^{2n})^2 (1+q^{2n-1})^4 \right\} . 
\label{eq:KK'}
\end{align} 
We can show 
\begin{align} 
k & =  4 q^{1/2} \prod_{n=1}^{\infty} \left\{ (1+q^{2n})/(1+ q^{2n-1}) \right\}^4 \, , \nn \\ 
k^{'} & = \prod_{n=1}^{\infty} \left\{ (1-q^{2n-1})/(1+ q^{2n+1}) \right\}^4 \, . 
\label{eq:kk'}
\end{align}

\subsubsection{Formula for the elliptic integral of the third kind}
By making transformation of variable $z=\sn(w)$ in eq. \eqref{ei3}, we have 
\begin{align}
\Pi(n,\phi=\am u) = \int_0^u \frac{\d w}{1-n\sn^2(w)} . 
\label{f3}
\end{align}
We define Jacobi's Zeta function $Z(u, k)$ by 
\begin{equation} 
Z(u, k ) = E(\am u, k) - u E/K . 
\end{equation}
We introduce Jacobi's Theta function by 
\begin{equation}
\Theta(u, k) = \Theta(0, k) \exp \left( \int_{0}^{u} Z(u, k) du \right) . 
\label{eq:df-Theta} 
\end{equation} 
In terms of $q=\exp(- \pi K^{'}/K)$ we have
\begin{equation} 
\Theta(u, k) = \prod_{n=1}^{\infty} \left\{(1-q^{2n})(1 - 2 q^{2n-1} \cos{\frac {\pi u} K} + q^{4n-2})  \right\} . 
\end{equation} 
By taking the logarithmic derivative of Jacobi's Theta function we have 
\begin{equation}
Z(u, k) = \frac {2 \pi} K \sin( \pi u/K) \sum_{n=1}^{\infty} 
\frac {q^{2n-1}} {1 - 2 q^{2n-1} \cos(\pi u/K) + q^{4n-2}} \, . 
\end{equation}

Let us take a complex number $a$ satisfying the following relation:  
\begin{equation} 
\sqrt n = k \, \sn(a, k) \, . 
\end{equation}
We can express the elliptic integral of the third kind in terms of Jacobi's Zeta function 
as follows. 
\begin{equation} 
\Pi(n, \phi=\am u, k) =  
u + \sqrt{\frac n {(k^2-n)(1-n)}} \left( 
{\frac 1 2} \log {\frac {\Theta(u-a, k)} {\Theta(u+a, k)}} + u \, Z(a, k) \right) \, . 
\label{eq:EI3}  
\end{equation} 
We express the complete elliptic integral of the third kind 
as follows. 
\begin{equation}
\Pi(n, k) = K \left( 1 + Z(a, k) \sqrt{\frac n {(k^2-n)(1-n)}} \right) \, .  
\label{eq:cei3} 
\end{equation}
\subsection{Jacobi's Eta function}
We define Jacobi's Eta function by 
\begin{equation} 
H(u,k) = 2 q^{1/4} \sin\left( \frac {\pi u} {2 K} \right) 
\prod_{n=1}^{\infty} \left\{ (1-q^{2n}) (1- 2 q^{2n} \cos( \pi u/K) + q^{4n} \right\} \, . 
\end{equation}
In terms of Eta and Theta functions Jacobi's $\sn$, $\cn$ and $\dn$ functions are given by 
\begin{align}
\sn u = \frac {H(u) \Theta (K)} {\Theta (u) H(K)} \, , \quad 
\cn u = \frac {H(u+K) \Theta (0)} {\Theta (u) H(K)} \, , \quad 
\dn u = \frac {\Theta(u+K) \Theta (0)} {\Theta(u) \Theta (K) } \, . 
\end{align} 
\subsection{Jacobi's imaginary transformation}
Jacobi's imaginary transformation for $\sn$, $\cn$ and $\dn$ functions 
is given by  \begin{align} 
 \sn(i u, k^{'}) & =  i \, \frac {\sn(u, k)} {\cn(u, k)} \, , \nn \\ 
 \cn(i u, k^{'}) & =  \frac 1 {\cn(u, k)} \, , \nn \\ 
 \dn(i u, k^{'}) & =  \frac {\dn(u, k)} {\cn(u, k)} \, . 
\end{align} 
Jacobi's imaginary transformation of the elliptic integral of the second kind is given by 
\begin{equation}
E(\am(iu), k) = u + \frac d {du} \log \cn(u, k^{'}) - i E(\am u, k^{'}) \, .  
\label{eq:ITE}
\end{equation}
We can derive Jacobi's imaginary transformation of Zeta function from \eqref{eq:ITE} 
\begin{equation} 
 Z(u, k) = i Z( i u, k^{'}) + \frac d {du} \log \cn(iu, k^{'}) - \frac {\pi u} {2 K K^{'}} \, . 
\end{equation}
It follows that we have 
\begin{align}
\frac {\Theta(u, k)} {\Theta(0, k)} 
  =  
\exp\left(- \frac {\pi u^2} {4 K K^{'}} \right) \frac {H(i u + K^{'}, k^{'})} {H(K^{'}, k^{'})} \, . 
\end{align} 
Therefore, in terms of $p=\exp(- \pi K/K^{'})$, we have  
\begin{equation} 
\frac {\Theta(u, k)} {\Theta(0, k)} = \exp\left(- \frac {\pi u^2} {4 K K^{'}} \right) 
\cosh\left( \frac {\pi u} {2 K^{'}} \right) \prod_{n=1}^{\infty} 
\frac { 1 + 2p^{2n} \cosh(\pi u/K^{'}) + p^{4n} }{ (1+p^{2n})^2 } \, . 
\label{eq:Theta-p} 
\end{equation}
By taking the logarithmic derivative 
of \eqref{eq:Theta-p} we expand Jacobi's Zeta function through \eqref{eq:df-Theta} 
as follows. 
\begin{align} 
Z(u, k) = - \frac {\pi u} {2K K^{'}} + 
\frac {\pi} {2 K^{'}} \tanh \left( \frac {\pi u} {2 K^{'}} \right) 
+ \frac {2 \pi} {K^{'}} \sinh \left( \frac {\pi u} {K^{'}} \right) 
\sum_{n=1}^{\infty} \frac {p^{2n}} {1 + 2 p^{2n} \cosh(\pi u/K^{'}) + p^{4n}} \, . 
\label{eq:Zeta-p}
\end{align} 

\setcounter{equation}{0} 
\renewcommand{\theequation}{B.\arabic{equation}}
\section{Gaussian weighted sum for the Dirichlet kernel}

In order to reduce the short-wavelength oscillations or ripples such as shown in the graph of the 
Dirichlet kernel  $D_{N_c}(x)$ we may modify the truncated sum \eqref{eq:truncatedsum}
with some weights.  Let us denote by $k$ the wavenumbers $k=2 \pi n/L$ for integers $n$.  
We take an integer $n_0$ and set  $k_0= 2 \pi n_0/L$ as a reference wavenumber. 
We define $\Delta n$ by $\Delta n= n- n_0$ and denote 
the difference of wavenumbers by $\Delta  k= k-k_0= 2 \pi \Delta n/L$.    
With the Gaussian weights $\exp(-\sigma^2 (k-k_0)^2)$
we modify the truncated sum \eqref{eq:truncatedsum}   as 
\begin{equation}
D_{N_c}^{\sigma} (x) =  \sum_{n=-N_c + n_0}^{N_c + n_0} \exp\left( i k x -\sigma^2 \Delta k^2 \right)  \label{eq:truncated_delta_Gaussian}
\end{equation}
In terms of $\Delta n$ we express it as 
\begin{equation}
D_{N_c}^{\sigma}(x) 
 =  \exp\left( - {\frac {x^2} {4 \sigma^2}} + i k_0 x \right) \sum_{\Delta n=-N_c}^{N_c} 
\exp \left(- \left(\sigma \Delta k - \frac {i x}{2 \sigma}\right)^2 \right)  \, .  
\label{eq:Gaussian_sum}
\end{equation}
We can show that the sum over $\Delta n$ in \eqref{eq:Gaussian_sum} does not become very large in the region of $x$ satisfying $|x|/\sigma < 4 \pi \sigma N_c/L$. If $|x|/\sigma > 4 \pi \sigma N_c/L$, each term in the sum \eqref{eq:Gaussian_sum} has its absolute value larger than 1, and hence the sum can be much larger than $2N_c+1$ which is the number of the terms in the sum. Here we remark that the bound $4 \pi \sigma N_c/L$ is given by 
$\pi$ if we set $\sigma =L/(4 N_c)$. We may therefore assume that 
 the sum over  $\Delta n$ in \eqref{eq:Gaussian_sum} is approximately constant 
in some region near the origin such as $|x| < 4 \sigma$. 
Evaluating the sum by an integral we approximate the Gaussian weighted sum 
$D_{N_c}^{\sigma}(x)$ as 
\begin{equation}
D_{N_c}^{\sigma}(x) = \frac L {2 \sqrt{\pi} \sigma} {\rm erf}(2 \pi \sigma N_c/L) 
\, \exp\left( - {\frac {x^2} {4 \sigma^2}} + i k_0 x \right)  \, .  
\end{equation}  
Here the error function ${\rm erf}(z)$ is given by 
\begin{equation}
{\rm erf}(z) = \frac 2 {\sqrt{\pi}} \int_{0}^{z} \exp(-x^2) dx \, . 
\end{equation}
Typically, by putting  $\sigma=L/(4N_c +2)$, i.e. half the period of  the short-distance oscillations 
or ripples in the graph of the function $D_{N_c}(x)$,  the ripples disappear, and the Gaussian weighted sum $D^{\sigma}_{N_c}(x)$ approximately becomes a smooth Gaussian wave packet with the standard deviation $2 \sigma$.


\end{document}